\documentclass[obeyspaces,
  journal=pasa,
  manuscript=research-paper,
  year=2025,
  volume=YY,
]{cup-journal}

\usepackage{graphicx} 
\usepackage{amsmath}
\usepackage{amssymb,microtype,siunitx,booktabs}
\usepackage[skip=0.5ex]{subcaption}
\usepackage{url}
\usepackage{hyperref} 

\def\fd{\hbox{$.\mkern-4mu^\circ$}}
\newcommand{\degr}{$^\circ$}
\newcommand{\arcmin}{$'$}
\newcommand{\arcsec}{$''$}
\newcommand{\zph}{$z_{\rm phot}$}
\newcommand{\zsp}{$z_{\rm spec}$}
\newcommand{\vsys}{$v_{\rm sys}$}
\newcommand{\kms}{~km\,s$^{-1}$}

\newcommand{\Msun}{~M$_{\odot}$}

\title{ASKAP Discoveries of Giant Radio Galaxies in the Sculptor field}

\author{B.S. Koribalski}
\affiliation{Australia Telescope National Facility, CSIRO, Space and Astronomy, P.O. Box 76, Epping, NSW 1710, Australia}
\alsoaffiliation{Western Sydney University, Locked Bag 1797, Penrith South DC, NSW 2751, Australia}
\email[B.S. Koribalski]{Baerbel.Koribalski@csiro.au}

\received {08 Apr 2025}
\revised  {dd Mmm 2025}
\accepted {dd Mmm 2025}
\published{dd Mmm 2025}


\begin{document}

\begin{abstract}
We present the discovery of 15 well-resolved giant radio galaxies (GRGs) with angular sizes $\ge$5~arcmin and physical sizes $>$1~Mpc in wide-field Phased Array Feed 944~MHz observations on the Australian Square Kilometre Array Pathfinder (ASKAP). We identify their host galaxies, examine their radio properties as well as their environment, and classify their morphologies as FR\,I (4), FR\,II (8), intermediate FR\,I/II (2), and hybrid (1). The combined $\sim$40 deg$^2$ ASKAP image of the Sculptor field, which is centred near the starburst galaxy NGC~253, has a resolution of 13\arcsec\ and an rms sensitivity of $\gtrsim$10~$\mu$Jy\,beam$^{-1}$. The largest GRGs in our sample are ASKAP J0057--2428 (\zph\ = 0.238), ASKAP J0059--2352 (\zph\ = 0.735) and ASKAP J0107--2347 (\zph\ = 0.312), for which we estimate linear projected sizes of 2.7, 3.5 and 3.8 Mpc, respectively. In total we catalog 232 extended radio galaxies of which 77 (33\%) are larger than 0.7 Mpc and 35 (15\%) are larger than 1~Mpc. The radio galaxy densities are 5.8 deg$^{-2}$ (total) and 0.9 (1.9) deg$^{-2}$ for those larger than 1 (0.7) Mpc, similar to previous results. Furthermore, we present the ASKAP discovery of a head-tail radio galaxy, a double-lobe radio galaxy with a spiral host, and radio emission from several galaxy clusters. As the ASKAP observations were originally conducted to search for a radio counterpart to the gravitational wave detection GW190814 ($z \sim 0.05$), we highlight possible host galaxies in our sample. 
\end{abstract}

\section{Introduction}
Giant radio galaxies are among the largest single objects in the Universe, typically defined as having projected linear sizes larger than 0.7~Mpc or 1~Mpc, depending on the study \citep[e.g.,][]{Schoenemakers2001,Kuzmicz2018,
Hardcastle2020, Dabhade2020, Andernach2021, Saikia2022, Simonte2022, Oei2023a}. This makes even the smallest GRGs $\sim$10--20 times larger than a typical Milky Way-like spiral galaxy and similar in size to the Local Group. GRGs give evidence to some of the most energetic processes inside their host elliptical galaxies and the morphologies of their radio lobes reflect the properties of their surrounding intergalactic medium (IGM). The presence of a host galaxy and their typically double-lobed radio morphology clearly distinguishes GRGs from other large radio sources such as cluster halos and cluster relics. 

Radio galaxies can be studied in great detail when well resolved by interferometric radio continuum observations. The large extent of giant radio lobes highlights their old age while their intricate shapes inform us about the local and large-scale environment, particularly density variations in the ambient IGM \citep[e.g.,][]{Malarecki2015, Peng2015}. During the active phase, the expanding jets and lobes forge a path through the IGM, while being impacted by the same medium. In contrast, during their inactive phase, the old radio lobes and their surrounding IGM slowly reach a pressure balance. In over half of the known GRGs, \citet{Bruni2019} find the central radio sources to be relatively young, likely linked to the episodic / re-starting activity of super-massive black holes (SMBHs); see also \citep{Jurlin2020}. 

Recent large-scale radio surveys such as the `Evolutionary Map of the Universe' \citep[EMU,][]{Norris2011, Norris2021-EMUPS} and the `Widefield ASKAP L-band Legacy All-sky Blind surveY' \citep[WALLABY,][]{Koribalski2012,Koribalski2020} projects, both conducted with the Australian Square Kilometre Array Pathfinder \citep[ASKAP,][]{Johnston2008, Hotan2021}, as well as the `LOFAR Two Metre Sky Survey' \citep[LoTSS,][]{Shimwell2019}, have resulted in many new discoveries and a resurgence of GRG studies. Both ASKAP's and LOFAR's large field of view, high resolution, dynamic range and good sensitivity to low-surface brightness structures have been essential to this research field, complemented by multi-colour optical sky surveys together with millions of photometric redshifts \citep[e.g.,][]{Bilicki2014, Bilicki2016, Zou2019, Zhou2021}. 

\begin{figure*}
\centering
 \includegraphics[width=17cm]{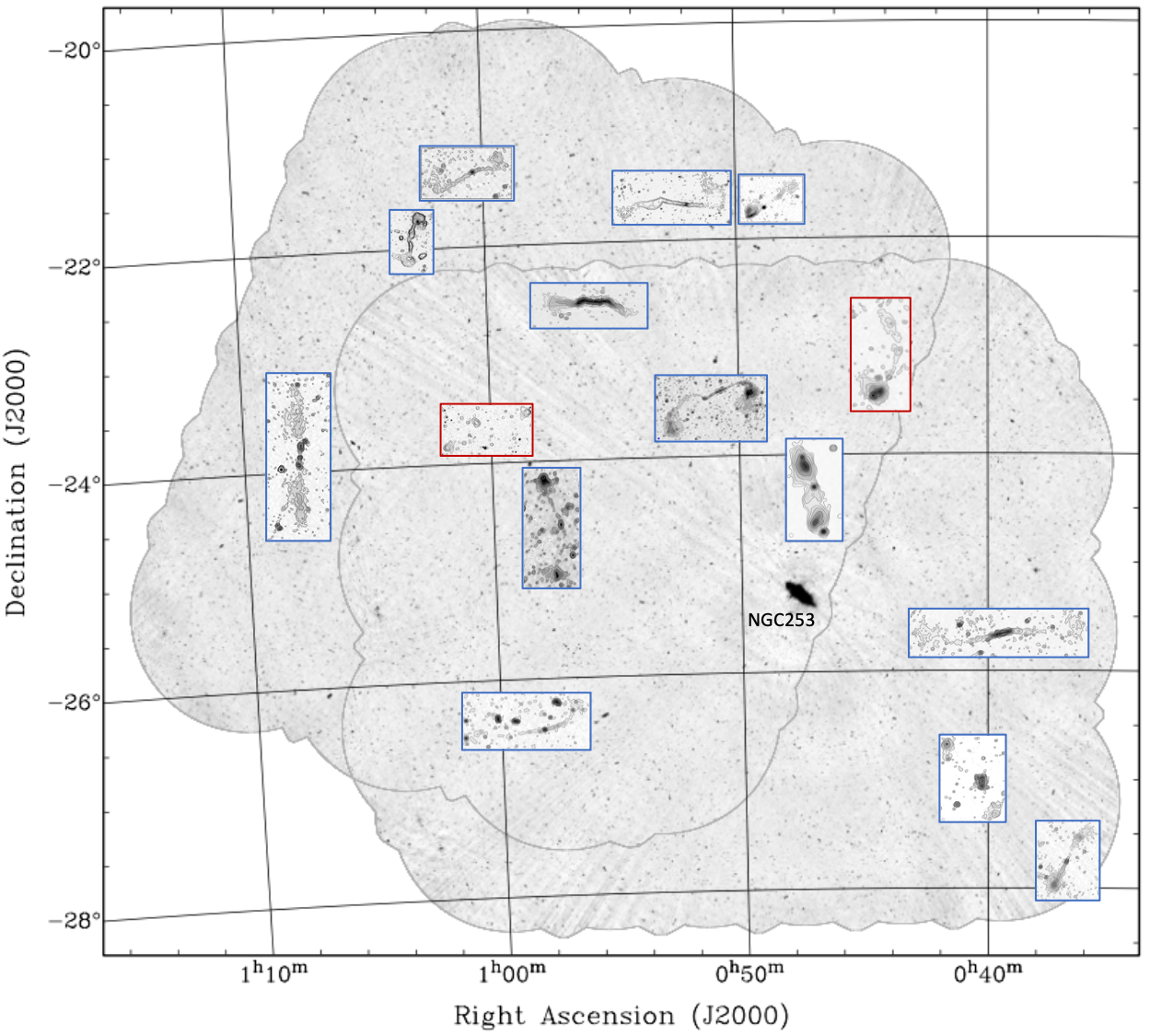}
\caption{Overview of the ASKAP 944~MHz Sculptor field (resolution 13 arcsec), consisting of a $7 \times 10$~h square field ($PA$ = 0\degr) and $1 \times 10$~h rotated square field ($PA$ = 67\fd5) offset the north-east. The field borders are indicated by grey lines; see also \citet[][their Fig.~1]{Dobie2022}. In the overlap region, which includes a large fraction of the GW190814 location area \citep{Abbott2020}, the average rms noise is $\sim$13 $\mu$Jy\,beam$^{-1}$. Residual artifacts from the bright starburst galaxy NGC~253 cause variations of the rms noise across the field. Overlaid are enlarged images of the 15 largest (in terms of angular size) giant radio galaxies in our sample listed in Table~1 (not to scale). The two candidate GRGs are indicted by red frames. }
\label{fig:overview-fig1}
\end{figure*}

In the wide-field ASKAP image of the Abell~3391/5 cluster (887.5~MHz, 30 deg$^2$, rms $\sim$30~$\mu$Jy\,beam$^{-1}$) \citet{Brueggen2021} found densities of 0.8 (1.7) deg$^{-2}$ for radio galaxies larger than 1 (0.7) Mpc, while \citet{Gurkan2022} found only 63 GRGs $>$0.7~Mpc in the ASKAP GAMA23 field (887.5~MHz, 83 deg$^2$, rms $\sim$38~$\mu$Jy\,beam$^{-1}$), i.e. $\sim$0.8 deg$^{-2}$. For the EMU Pilot Survey (944~MHz, 270 deg$^2$, rms $\sim$25--30~$\mu$Jy\,beam$^{-1}$) \citet{Norris2021-EMUPS} report a preliminary number of at least 120 GRGs $>$1 Mpc (and a similar number with sizes between 0.7 and 1~Mpc) among the $\sim$220\,000 catalogued sources. \citet{Andernach2021} present the discovery of 178 GRGs $>$1~Mpc within 1059 deg$^2$, a small area within the shallow Rapid ASKAP Continuum Survey \citep[RACS,][]{McConnell2020} at 887.5~MHz (RACS-low, DEC $<$ +40 deg, rms $\sim$250 $\mu$Jy\,beam$^{-1}$). In the LOTSS Bo\"otes deep field at 150~MHz (rms $\sim$30~$\mu$Jy\,beam$^{-1}$) \citet{Simonte2022} find somewhat higher sky densities of $\sim$1.4 (2.8) deg$^{-2}$ GRGs with linear sizes $>$1 (0.7) Mpc. As the survey depth, frequency and angular resolution vary substantially, these numbers are only indicative and likely lower limits. \\

\begin{figure*}
 \centering
 \includegraphics[height=8cm]{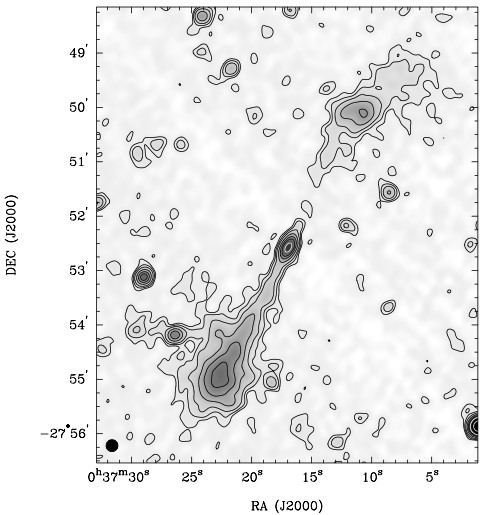}
 \includegraphics[height=8cm]{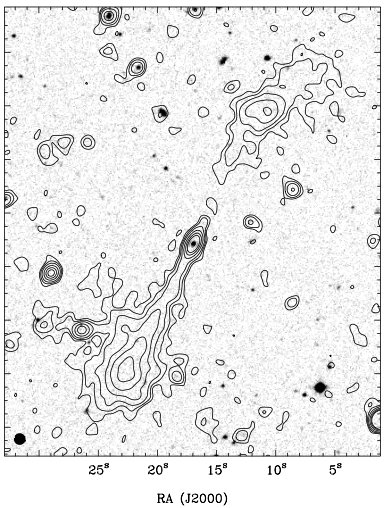} \\
 \caption{ASKAP J0037--2752 (FR\,II-type GRG). --- {\bf Left}: ASKAP 944~MHz radio continuum map; the contour levels are 0.04, 0.1, 0.2, 0.4, 0.65, 0.9, 1.3, 3 and 5 mJy\,beam$^{-1}$. --- {\bf Right}: ASKAP radio contours overlaid onto a DSS2 $R$-band image. The GRG host galaxy is WISEA~J003716.97--275235.3 (\zsp\ = 0.2389). The ASKAP resolution of 13 arcsec is shown in the bottom left corner.} 
 \label{fig:J0037-2752}
\end{figure*}

\begin{figure*}
 \centering
 \includegraphics[width=14cm]{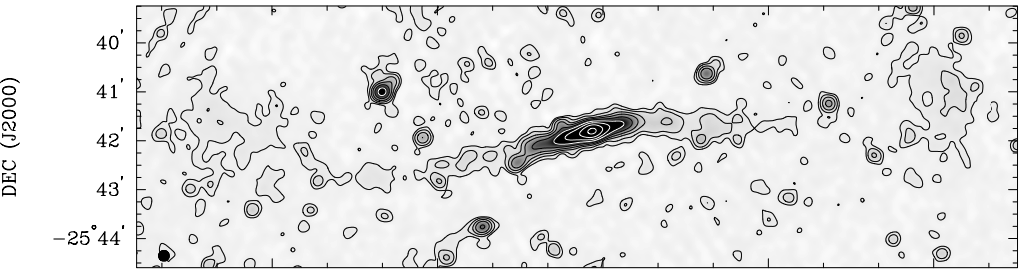} \\
 \includegraphics[width=14cm]{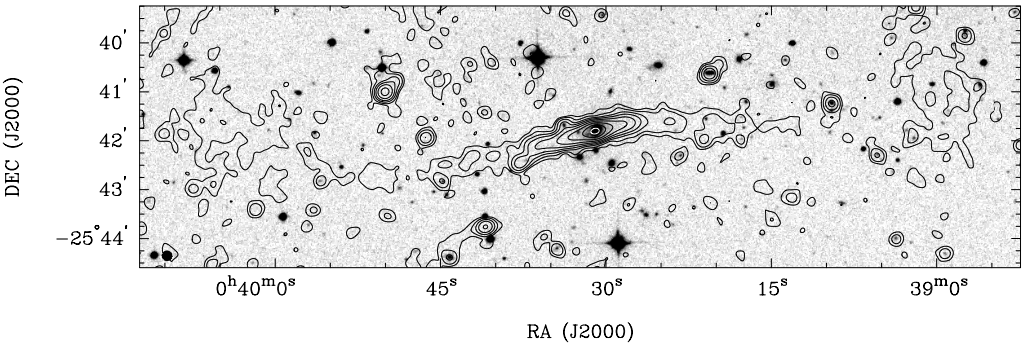} \\
 \caption{ASKAP J0039--2541 (FR\,I-type GRG). --- {\bf Top}: ASKAP 944~MHz radio continuum map; the contour levels are 0.03, 0.1, 0.25, 0.5, 1, 2, 4, 8 and 16 mJy\,beam$^{-1}$. --- {\bf Bottom}: ASKAP radio contours overlaid onto a DSS2 $R$-band image. The GRG host galaxy is WISEA~J003930.86--254147.8 (\zsp\ = 0.073). The ASKAP resolution of 13 arcsec is shown in the bottom left corner.}
 \label{fig:J0039-2541}
\end{figure*}

Radio galaxies (RGs) come in a wide range of morphologies \citep[e.g.,][]{Banfield2015}, the most common of which are briefly described below. We note that RG classifications can change when more detailed (higher sensitivity / resolution) images become available.
\begin{itemize}
    \item Fanaroff-Riley Class~I (FR\,I) galaxies have bright inner radio jets and fading outer radio lobes without hotspots (edge-darkened). Typical examples of this class are 3C\,449 \citep{Feretti1999} and IC\,4296 \citep{Condon2021}.
    \item Fanaroff-Riley Class~II (FR\,II) galaxies are characterised by prominent radio hot spots at the end of their radio lobes (edge-brightened). A typical example of this class is 3C\,98 \citep{Leahy1997}. 
    \item Hybrid Morphology Radio Sources (HyMoRS) show a mix of FR\,I and FR\,II morphology \citep[e.g.,][]{Harwood2020,Stroe2022}. Some radio galaxies, like Hercules\,A (3C\,348) are classified as intermediate FR\,I/II sources \citep[e.g.,][]{Timmerman2022}.
    \item X-shaped radio galaxies (XRG), like the GRG PKS\,2014--55, consist of a double-lobed radio galaxy plus a set of secondary lobes, referred to as wings, likely due to backflow \citep[e.g.,][]{Cotton2020}.
    \item The lobes of radio galaxies are shaped by their surrounding intergalactic medium (IGM) -- esp. once the jets have turned off -- and therefore display a wide variety of shapes. So-called "bent tail" (BT) galaxies are sometimes classified as wide-angle tail (WAT) or narrow-angle tail (NAT) radio galaxies depending on the jet opening angle. But the tail appearances (and classifications) can vary hugely with image resolution and sensitivity, as shown for example by \citet{Gendron-Marsolais2020} for NGC~1265. A subset of head-tail (HT) radio galaxies have two highly-bent inner jets forming a single tail \citep[e.g., the Corkscrew Galaxy,][] {JonesMcAdam1996,Koribalski2024-Corkscrew}. Bent tail radio galaxies are often (but not always) found in clusters \citep[e.g.,][]{Veronica2022, Ramatsoku2020}.
    \item Remnant radio galaxies typically have two diffuse (amorphous), low surface brightness (LSB) radio lobes, and a weak radio core. They have neither jets nor hot spots, and their fading lobes are recognizable by steep spectral indices \citep[e.g.,][]{Cordey1987,Tamhane2015,Brienza2016,Randria2020}. In these galaxies the central SMBH has been inactive for some time and will likely re-start when triggered \citep{Jurlin2020,Shabala2020}.
    \item Double-double radio galaxies (DDRG) have two sets of double lobes, typically one outer set of remnant (old) lobes and one inner set of new (young / re-started) radio lobes \citep[e.g.,][]{Saripalli2012,Kuzmicz2017}.
\end{itemize}

In this paper we focus on the 15 newly-discovered GRGs in the ASKAP Sculptor field with largest angular sizes (LAS) $\ge$5 arcmin and projected largest linear sizes (LLS) $>$1~Mpc, whose properties are summarised in Tables~1--3. The projected linear sizes are lower limits to their actual sizes, as neither inclination nor curvature is taken into account. Furthermore, deeper radio images often reveal larger sizes, esp. when the lobe emission is of very low surface brightness. The full sample of catalogued RGs in the field is presented in the Appendix. -- We adopt the following cosmological parameters: $H_{\rm 0}$ = 70 km\,s$^{-1}$\,Mpc$^{-1}$, $\Omega_{\rm m}$ = 0.3, and $\Omega_{\rm \Lambda}$ = 0.7.

\begin{table*}
\centering
\begin{tabular}{cccccccc}
\hline
 ASKAP & host name / position & \multicolumn{2}{c}{redshift} & \multicolumn{2}{c}{extent of radio lobes} \\
 Name & $\alpha,\delta$ (J2000) & \zsp & \zph & LAS & LLS & type & comments \\
 & [hms, dms] & & & [arcmin] & [Mpc] \\
\hline
\hline
J0037--2752 & WISEA J003716.97--275235.3 & 0.23887 & 0.233 & 7.5 & 1.70 & FR\,II & twin jets \\
J0039--2541 & WISEA J003930.86--254147.8 & 0.07297 & 0.069 & 15.5 & 1.29 & FR\,I & relic lobes \\
J0041--2655 & WISEA J004119.25--265548.3 & -- & 0.232 & 7.3 & 1.62 & FR\,II & relic lobes \\
J0044--2317 & WISEA J004426.72--231745.8 & -- & 0.362 & 6.0? & 1.82 & HyMoRS & GRG candidate, asym. \\
J0047--2419 & WISEA J004709.94--241939.6 & -- & 0.270 & 5.0 & 1.24 & FR\,II & relic lobes \\ \\
J0049--2137 & WISEA J004941.58--213722.1 & -- & 0.233 & 7.2 & 1.59 & FR\,II & remnant, asym.  \\
J0050--2135 & WISEA J005046.49--213513.6 & 0.05760 & 0.056 & 18.0 & 1.20 & FR\,I & asym. \\ 
J0050--2325 & WISEA J005049.89--232511.1 & 0.11137 & 0.110 & 13.5 & 1.60 & FR\,II, WAT & twin jets \\
J0055--2231 & WISEA J005548.98--223116.9 & 0.11437 & 0.116 & 9.2  & 1.14 & FR\,I \\
J0057--2428 & WISEA J005736.30--242814.9 & -- & 0.238 & 12.1 & 2.74 & FR\,II & inner radio knots \\ \\
J0058--2625 & WISEA J005835.74--262521.3 & 0.11341 & 0.118 & 12.0 & 1.48 & FR\,I/II, WAT? & LEDA 3237521, twin jet \\ 
J0059--2352 & WISEA J005954.72--235254.7 & -- & 0.735 & 8.0 & 3.49 & FR\,II & GRG candidate \\
J0100--2125 & WISEA J010039.00--212533.5 & -- & 0.193 & 6.3 & 1.21 & FR\,I  \\
J0102--2154 & WISEA J010245.22--215414.3 & 0.2930 & 0.284 & 6.4 & 1.68 & FR\,I/II & relic lobes, precessing \\
J0107--2347 & WISEA J010721.41--234734.1 & -- & 0.312 & 13.8 & 3.79 & FR\,II, DDRG & re-started and remnant lobes \\
\hline
\end{tabular}
\caption{Properties of the 15 GRGs with the largest angular sizes in the ASKAP Sculptor field and their respective host galaxies. In Col.~(2) we chose the WISE names of the host galaxies, while each has numerous designations. Spectroscopic redshifts (\zsp) were obtained from 2dF \citep{Colless2001} or 6dF \citep{Jones2009} as noted in Section~3.1. Photometric redshifts (\zph) were obtained from DES-DR9 \citep{Zhou2021}.}
\label{tab:GRG-tab1}
\end{table*}

\section{ASKAP Observations and Data Processing}
ASKAP is a 6~km diameter radio interferometer consisting of $36 \times 12$-m antennas, each equipped with a wide-field Phased Array Feed (PAF), and operating at frequencies from 700 MHz to 1.8 GHz \citep{Johnston2008}. The currently available correlator bandwidth of 288~MHz is divided into $288 \times 1$ MHz coarse channels; the typical field-of-view is 30 deg$^2$. For a comprehensive overview see \citet{Hotan2021}. ASKAP science highlights are presented in \citet{Koribalski2022-IEEE}.

We obtained nine fully processed ASKAP radio continuum images from the CSIRO ASKAP Science Data Archive (CASDA), observed between Aug 2019 and Dec 2020 with the band centred at 944 MHz. The ASKAP PAFs were used to form 36 beams arranged in a closepack36 formation, each delivering $\sim$30 deg$^2$ field of view. All but one of the ASKAP fields were observed for $\sim$10~h and have an average rms noise of $\sim$37 $\mu$Jy\,beam$^{-1}$. When combining the ASKAP images, we omitted the short-integration (3.5~h) field due to its larger beam size. Seven fields have the same pointing centre, while the eighth field is slightly offset to the north-east and rotated by 67.5\degr. Figure~1 shows the combined area of $\sim$40 deg$^2$. These are the same data used to analyse ORC~J0102--2450 \citep{Koribalski2021}. The field centres are close to the nearby  (\vsys\ = $243 \pm 2$\kms) starburst galaxy NGC~253 \citep{Koribalski1995, Koribalski2004}, which resides in the Sculptor Group. The radio brightness and large extent of the NGC~253 disc cause minor artifacts over part of the field.  
For a summary of the ASKAP observations, which were conducted to search for the radio counterpart of the gravitational wave event GW190814 \citep{Abbott2020}\footnote{GW190814 was detected on the 14th of August 2019 at 21:10:39 UTC by the LIGO-VIRGO Consortium (LVC). Its localisation area is 18.5 deg$^2$ at 90\% probability, with the larger of the two areas centred near  $\alpha,\delta$(J2000) $\sim 00^{\rm h}\,51^{\rm m}$, --25\degr\ just north-east of the foreground starburst galaxy NGC~253. Modelling of GW190814 suggests it is coalescing binary consisting of a 23\Msun\ black hole and a 2.6\Msun\ compact object, located at a distance of 196 -- 282~Mpc or $z \sim 0.05$ \citep{Abbott2020}. The compact object could be a neutron star (NS) or a black hole (BH).}, see \citet{Dobie2022}. The data processing was done with the ASKAPsoft pipeline \citep{Whiting2017,Wieringa2020}. We combined all eight $\sim$10~h integration images after convolving each to a common 13\arcsec\ resolution, achieving an rms sensitivity of $\sim$13 $\mu$Jy\,beam$^{-1}$ in the artifact-free parts of the overlap region. 

\begin{figure}
\centering
 \includegraphics[height=5.3cm]{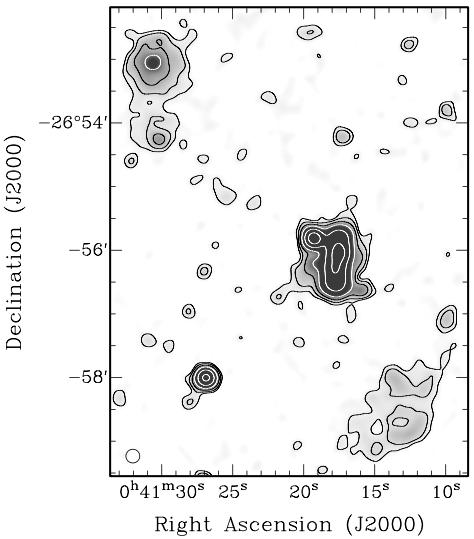}
 \includegraphics[height=5.3cm]{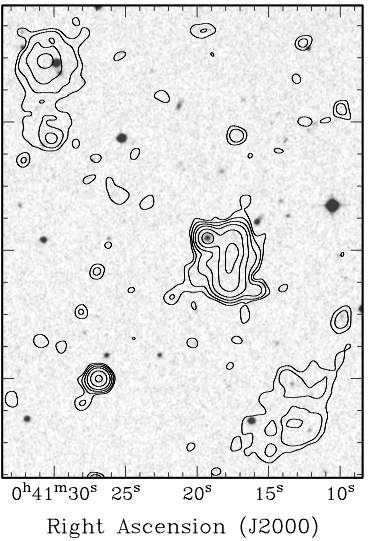}
\caption{ASKAP J0041--2655 (FR\,II-type GRG). --- {\bf Left:} ASKAP 944 MHz radio continuum map; the contour levels are 0.06, 0.12, 0.25 mJy\,beam$^{-1}$ (black), and 0.5, 1, 1.2, and 2.5 mJy\,beam$^{-1}$ (white). The ASKAP resolution of 13 arcsec is shown in the bottom left corner. --- {\bf Right:} ASKAP radio contours overlaid onto a DSS $R$-band optical image. The GRG host galaxy is WISEA J004119.25--265548.3 (\zph\ = 0.232). --- Superimposed is another double-lobe radio galaxy (ASKAP J0041--2656, LAS $\sim$ 1 arcmin; \zph\ = 0.713), located south-west of the radio core of the GRG ASKAP J0041--2655.}
 \label{fig:J0041-2655}
\end{figure}

\begin{figure}
 \centering
 \includegraphics[height=6.5cm]{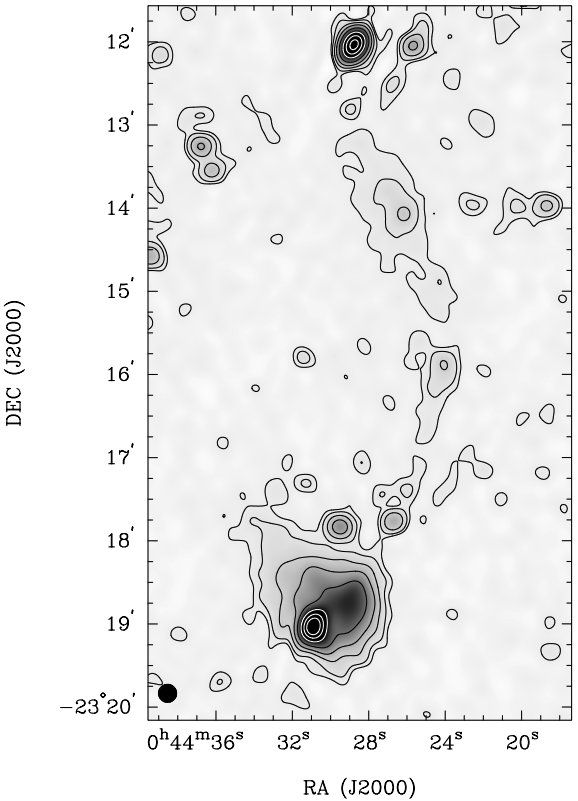}
 \includegraphics[height=6.5cm]{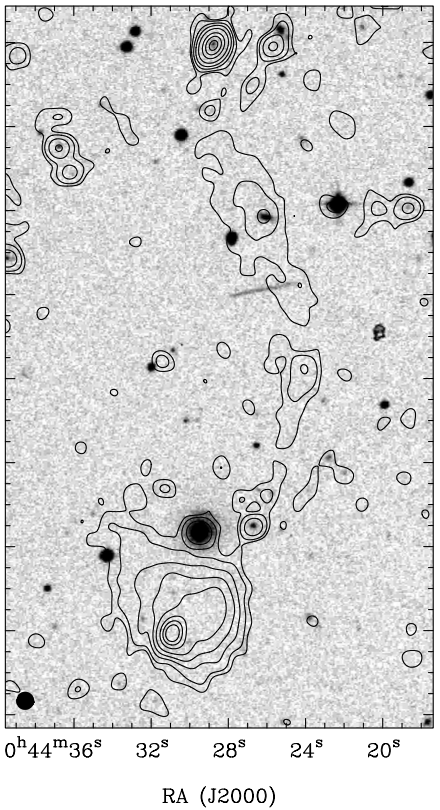}
\caption{ASKAP J0044--2317 (highly asymmetric HyMoRS-type GRG candidate). --- {\bf Left}: ASKAP 944~MHz radio continuum map; the contour levels are 0.04, 0.1, 0.25, 0.5, 1, 2, 3 and 4 mJy\,beam$^{-1}$. The ASKAP resolution of 13 arcsec is shown in the bottom left corner. --- {\bf Right}: ASKAP radio contours overlaid onto a DSS2 $R$-band image. The likely GRG host galaxy is WISEA~J004426.72--231745.8 (\zph\ = 0.362).  --- The prominent foreground spiral galaxy WISEA~J004429.50--231749.7 (\zsp\ = 0.060), located just north of the southern lobe and east of the GRG host galaxy, is detected with 0.74 mJy.}
 \label{fig:J0044-2317}
\end{figure}
 
\begin{figure} 
 \centering
 \includegraphics[height=7cm]{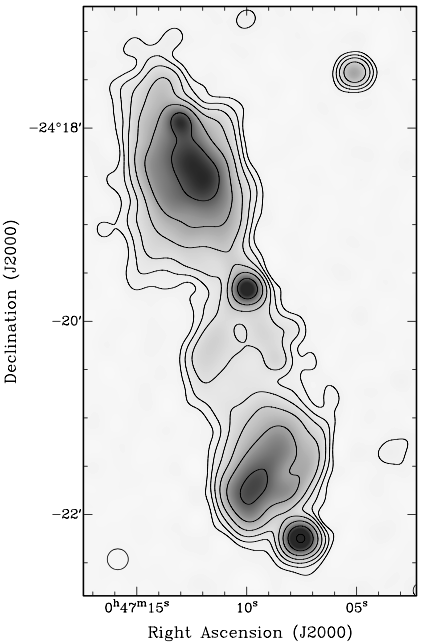}
 \includegraphics[height=7cm]{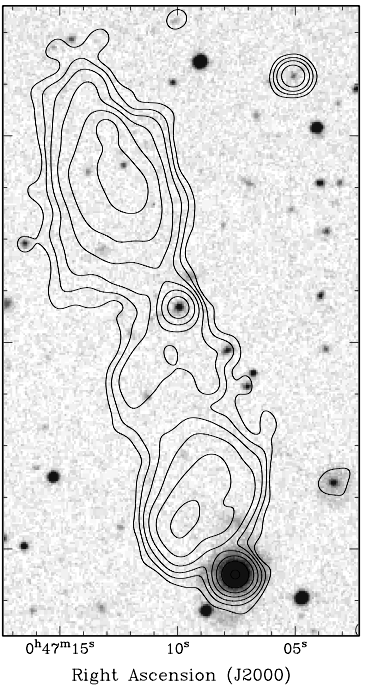}
\caption{ASKAP J0047--2419 (FR\,II-type GRG). --- {\bf Left}: ASKAP 944~MHz radio continuum map; the contour levels are 0.1, 0.25, 0.5, 1, 2, and 4 mJy\,beam$^{-1}$. The ASKAP resolution of 13 arcsec is shown in the bottom left corner. --- {\bf Right}: ASKAP radio contours overlaid onto a DSS2 $R$-band image. The GRG host galaxy is WISEA~J004709.94--241939.6 (\zph\ = 0.270). --- Just south of the extended radio lobes we detect a $\sim$10 mJy radio source coincident with the merging galaxy system ESO\,474-G026 (\zsp\ = 0.05271). The face-on star-forming spiral LEDA~790836 (\zph\ $\sim$ 0.08), located just west of the southern lobe, is also detected ($\sim$0.3 mJy). A DES-DR10 optical image of both galaxies is shown in Fig.~7, and more details are given in Section~3.1.}
 \label{fig:J0047-2419}
\end{figure}

\begin{figure}
\centering
 \includegraphics[width=8cm]{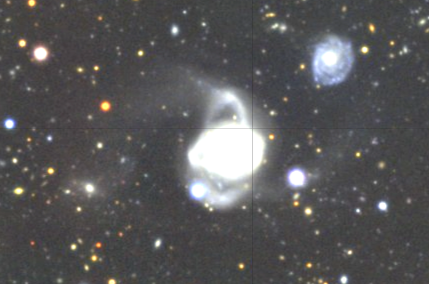}
\caption{DES-DR10 optical colour image of the galaxies ESO\,474-G026 (centre) and LEDA~790836 (top right). The contrast is chosen to show the newly discovered, very faint stellar tails extending to the east and west of the merging galaxy system ESO\,474-G026. The ASKAP radio continuum emission of both galaxies is evident in Fig.~6 which is centred on the FR\,II-type GRG ASKAP J0047--2419.}
\label{fig:eso474}
\end{figure}

\section{Results}

Figure~1 shows the $\sim$40 deg$^2$ ASKAP field studied here. It consists of a deep field ($\sim$70~h) covering $00^{\rm h}\,37^{\rm m} <$ $\alpha$(J2000) $< 01^{\rm h}\,04^{\rm m}$ and --22\degr\,25\arcmin\ $<$ $\delta$(J2000) $<$ --28\degr\,10\arcmin, and a rotated ($\sim$10~h) field offset to the north-east. Our analysis of the ASKAP data is complemented by optical, infrared, X-ray and other radio data. Specifically, we make use of the deep multi-band optical images from the Dark Energy Surveys \citep[DES,][]{DES2016} as well as radio continuum images from the 2--4~GHz Very Large Array Sky Survey \citep[VLASS,][]{Lacy2020}, the 1.4~GHz NRAO VLA Sky Survey \citep[NVSS,][]{Condon1998}, the 150~MHz TIFR GMRT Sky Survey  \citep[TGSS,][]{Intema2017}, and the 72--231~MHz GaLactic and Extragalactic All-sky Murchison Widefield Array survey \citep[GLEAM,][]{Hurley2017, Hurley2022}. \\

We conducted a by eye search for radio galaxies with large angular sizes, similar to \citet[][for NVSS]{Lara2001} and \citet[][for SUMSS]{Saripalli2005}. While our primary focus was on radio structures larger than $\sim$5~arcmin, we tried to catalogue all radio sources larger than $\sim$1~arcmin. Due to some imaging artifacts in the field, especially around NGC~253 and other bright radio sources, we did not use any source finding tools. No GRGs were catalogued by \citet{Kuzmicz2018} in this field, and, apart from the nearby, starburst galaxy NGC~253 \citep{Koribalski1995,Koribalski2018}, only one bright radio galaxy (NVSS J003757--250425) was catalogued by \citet{vanVelzen2012}. We paid particular attention to large LSB structures such as diffuse, extended radio lobes, cluster halos and relics, as well as odd radio circles. \\

We catalogue 35 giant radio galaxies with LLS $>$ 1~Mpc, incl. four candidates. Furthermore, we catalogue 42 RGs with $0.7 <$ LLS $< 1.0$~Mpc, incl. three candidates, and 155 RGs with LLS $< 0.7$~Mpc. These numbers suggest a source density of $\sim$0.9 deg$^{-2}$ for GRGs $>$1~Mpc and $\sim$1.9 deg$^{-2}$ for those $>$0.7~Mpc. In total, we catalogue 232 RGs (listed in the Appendix), of which 164 (70\%) are classified as FR\,II, 30 are classified as FR\,I/II, 29 as FR\,I and nine others.

\subsection{GRGs with large angular sizes}
In the following we discuss in detail the 15 GRGs with the largest angular sizes (LAS $\ge$ 5 arcmin; see Table~1) in the ASKAP Sculptor field. Fig.~2--18 show ASKAP images as well as optical $R$-band images from the Digitized Sky Survey (DSS2), both overlaid with radio contours, as well as a few zoom-in images. For each GRG we list the most likely host galaxy together with its redshift, the GRG's projected angular size, its linear size, and its morphological type. Fig.~19 shows DES-DR10 optical images of the GRG host galaxies as well as four others. The largest angular size (LAS) of a radio galaxy is measured along a straight line connecting opposite "ends" of the radio source. For FR\,II sources, we measure the LAS between the centres of the two hotspots. Only for very faint / diffuse lobes do we measure LAS out to the 3$\sigma$ contour. For bent-tailed sources, we measure the LAS along a straight line between the most separate diametrically opposite emission regions. Because of projection effects, bending of the tails, and surface-brightness sensitivity, the stated GRG extent is nearly always a lower limit. The WISE magnitudes and colours of the GRG hosts are given in Table~2 and the ASKAP 944 MHz total and component flux densities are listed in Table~3. \\

{\bf ASKAP~J0037--2752} is an FR\,II-type GRG with a bright core and two extended lobes (LAS = 7.5 arcmin; $PA \sim 150^\circ$, see Fig.~2). The southern radio lobe, which connects to the core, is significantly brighter than the disconnected, more diffuse northern lobe. The radio core is associated with the galaxy WISEA~J003716.97--275235.3 (2MASX J00371697--2752350, LEDA~3199033; DES J003716.97--275235.4) at \zsp\ = 0.23887 \citep[2dF,][]{Colless2001}. We estimate a linear extent of 1.70~Mpc for the system. 
The GRG radio core was previously catalogued as NVSS~J003717--275242 ($2.9 \pm 0.6$ mJy at 1.4~GHz); it is also detected in VLASS. 
We measure an ASKAP core position of $\alpha,\delta$(J2000) = $00^{\rm h}\,37^{\rm m}\,16.95^{\rm s}$, --27\degr\,52\arcmin\,34.8\arcsec, a 944~MHz peak flux of 1.5 mJy\,beam$^{-1}$ and an integrated 944~MHz flux of 2.8 mJy. It is likely the latter value includes radio emission from inner jets. The GRG's northern (N) and southern (S) lobes were previously catalogued as NVSS~J003712--275003 ($15.7 \pm 3.7$ mJy) and NVSS~J003722--275445 ($14.7 \pm 3.6$ mJy), respectively. We obtain ASKAP 944~MHz integrated flux densities of 6.7 mJy (N), 18.7 mJy (S) and 28.2 mJy (total). \\

{\bf ASKAP J0039--2541} is an FR\,I-type GRG with a bright core, inner jets and very faint relic lobes spanning 15.5 arcmin from east to west (see Fig.~3). Its radio core is associated with the galaxy WISEA~J003930.86--254147.8 (2MASX J00393086--2541483, DES J003930.85--254147.8) at \zsp\ = 0.07297 \citep[2dF,][]{Colless2001}.
We estimate a linear extent of 1.29 Mpc for the system. The GRG radio core was previously catalogued as NVSS J003931--254149 ($40.1 \pm 1.6$ mJy at 1.4~GHz); it is also detected in TGSS and VLASS. We measure an ASKAP core position of $\alpha,\delta$(J2000) = $00^{\rm h}\,39^{\rm m}\,31.0^{\rm s}$, --25\degr\,41\arcmin\,48.6\arcsec, a 944~MHz peak flux of 19.1 mJy\,beam$^{-1}$ and a total integrated 944~MHz flux of 68.3 mJy. \\

{\bf ASKAP J0041--2655} is an FR\,II-type GRG with a radio core and two relic lobes (LAS = 7.3 arcmin, see Fig.~4). Its likely host galaxy is WISEA J004119.25--265548.3 (DES J004119.25--265548.1) with \zph\ = 0.232, suggesting LLS = 1.62 Mpc. Just south-west of its radio core is another, much smaller double-lobed radio galaxy (ASKAP J0041--2656, also catalogued as NVSS J004118--265603) with LAS $\sim$ 1 arcmin and host galaxy WISEA J004118.07--265601.8 (\zph\ = 0.713); we derive LLS = 430 kpc. We find no X-ray emission that would hint at a cluster environment. We measure the following flux densities for ASKAP J0041--2655: $\sim$4 mJy (core), 5.3 mJy (N), 3.7 mJy (S), and 13 mJy (total). The radio core is clearly detected in VLASS, showing a possible N--S extension.  \\

{\bf ASKAP J0044--2317} is a GRG candidate with LAS $\sim$ 6.0 arcmin (see Fig.~5). Its near circular southern lobe (S) has a bright hotspot, while its northern lobe (N) is long, narrow and bent (extending to $\delta$ = -23\degr\,13\arcmin\,15\arcsec), resulting in a very asymmetric appearance. We suggest it is a HyMoRS candidate with likely host galaxy WISEA~J004426.72--231745.8 (DES~J004426.71--231745.7). Based on \zph\ = 0.362 we derive LLS = 1.82 Mpc. 
We measure flux densities of 26.0 mJy (S), 3.5 mJy (N), 0.5 mJy (core) and 30.0 mJy (total). The southern lobe was already catalogued as NVSS J004429--231829, but is resolved out in VLASS. The nearest known cluster is WHL J004347.8--231714
at \zph\ = 0.395 \citep{WHL2012}.

A faint foreground galaxy is coincident with the brightest part of the northern-most lobe ($\delta$ = -23\degr\,14\arcmin, \zph\ $\sim$ 0.15). Another prominent foreground spiral \citep[\zsp\ = 0.060,][]{Jones2009} is detected just north of the southern lobe. \\

{\bf ASKAP J0047--2419} is an FR\,II-type GRG with bright radio lobes extending over 5.0 arcmin (see Fig.~6) corresponding to LLS = 1.24 Mpc at the adopted host galaxy (WISEA J004709.94--241939.6, DES J004709.93--241939.5) redshift of \zph\ = $0.270 \pm 0.038$ \citep{Zou2019}.
Faint optical tails are detected in DES-DR10 around the host galaxy, most prominent to the south-east. The galaxy's extreme WISE colours led \citet{Flesch2015} to consider it a quasar at $z \sim 0.3$; they also note an associated X-ray source XMMSL J004710.0--241939.
The ASKAP flux measurements are listed in Table~2. VLASS 3~GHz images show a hint of radio emission from inner jets, extended approx. N--S. The radio source is also detected in TGSS at 150~MHz, and NVSS--TGSS spectral index maps are available, showing $\alpha = -0.8 \pm 0.3$. See \citet{Spavone2012} for an NVSS image of the GRG. \\

South of ASKAP J0047--2419, we detect a $\sim$10 mJy radio source coincident with the merging galaxy system ESO\,474-G026 \citep[\zsp\ = 0.05271,][]{Galletta1997}. The deep DES-DR10 optical image (see Fig.~7) highlights the merger's spectacular stellar rings \citep{Reshetnikov2005, Spavone2012} as well as two previously unknown broad tails of extremely low-surface brightness curving to the east and west, together spanning $\sim$3 arcmin. \citet{Reshetnikov2005} derive an H\,{\sc i} mass of $2 \times10^{10}$\Msun\ and a star-formation rate of 43\Msun\,yr$^{-1}$ for ESO\,474-G026. High-resolution VLASS images reveal a radio core plus faint bi-polar jets aligned approx. N--S, hinting at a central active galactic nucleus (AGN). Within the uncertainties, the two VLASS epochs show the same source morphology and flux densities. 
ESO\,474-G026's location and redshift make it a possible host of GW190814. Major merger systems like ESO\,474-G026 have much increased star formation rates compared to isolated galaxies \citep{Mihos1996, Hopkins2013, Moreno2021}, and contain large numbers of young star clusters which are ideal locations for BH--BH and BH--NS mergers \citep[e.g.,][]{Ziosi2014, DiCarlo2020, Mandel2022}. 
As a consequence, stellar-mass mergers detected by LIGO are more likely to occur in massive, merging galaxies than isolated galaxies. Since \citet{Dobie2021} find no radio afterglow in the ASKAP data, they suggest ESO\,474-G026 is unlikely the GW190814 counterpart. \\

{\bf ASKAP J0049--2137} is an FR\,II-type remnant GRG with host galaxy WISEA J004941.58--213722.1 (\zph\ = 0.233) and LAS = 7.2 arcmin, suggesting LLS = 1.59 Mpc. The GRG has a very asymmetric appearance (see Fig.~8). Its bright SE lobe extends $\sim$2.5 arcmin from the compact core and it appears to be bent backwards, while the NW lobe is fainter and extends nearly 5 arcmin. We measure ASKAP flux densities of 7 mJy (core), 38 mJy (SE lobe), 16 mJy (NW lobe), and 61 mJy (total). The core is detected in VLASS and NVSS images, while the SE lobe is only seen in NVSS. \\

\begin{figure} 
 \centering
 \includegraphics[width=8cm]{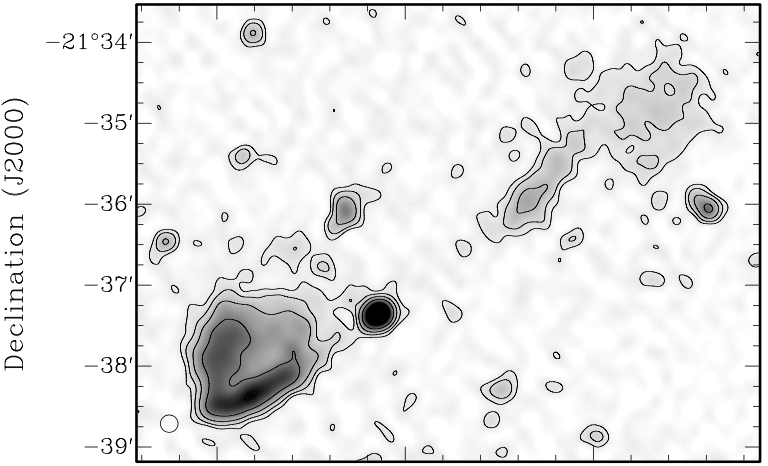}
 \includegraphics[width=8cm]{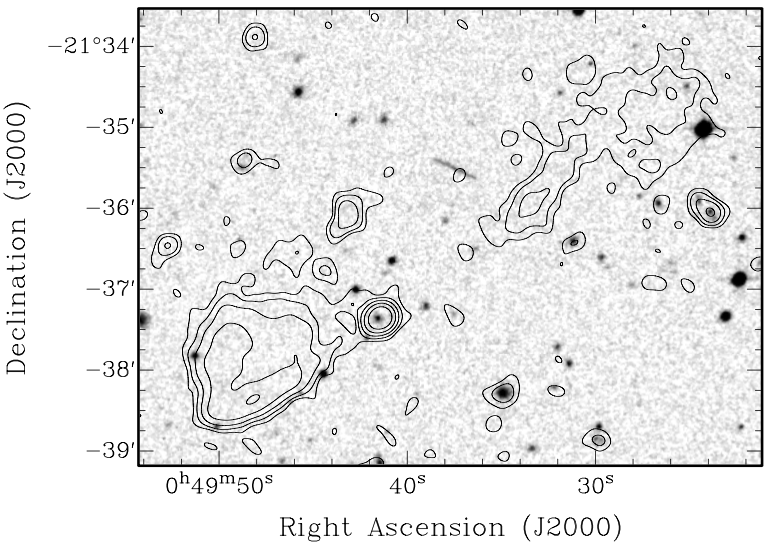}
\caption{ASKAP J0049--2137 (FR\,II-type GRG). -- {\bf Top}: ASKAP 944~MHz radio continuum map; the contour levels are 0.12, 0.25, 0.5, 1 and 2 mJy\,beam$^{-1}$. --- {\bf Bottom}: ASKAP radio contours overlaid onto a DSS2 $R$-band image. The GRG host galaxy is WISEA~J004941.58--213722.1 (\zph\ = 0.233). The ASKAP resolution of 13 arcsec is shown in the bottom left.}
 \label{fig:J0049-2137}
\end{figure}

{\bf ASKAP J0050--2135} is an asymmetric FR\,I-type GRG with LAS = 18 arcmin; see Fig.~9. Bipolar jets emerge from its elliptical host galaxy, WISEA J005046.49--213513.6 \citep[\zsp\ = 0.05760,][]{Jones2009}. The eastern jet/tail is much brighter and longer, at least in projection, than the western jet/tail. It bends ($\sim$45\degr) and broadens after $\sim$4 arcmin, with remarkably sharp boundaries, before ending in a faint patch of emission. A similar kink is also seen in the Barbell GRG \citep{Dabhade2022}. The short western jet of ASKAP J0050--2135 fades after 2--3 arcmin and curves to the north, then back east to form a hook. The GRG's LLS is 1.20~Mpc. It is located between two clusters, Abell~114 and Abell~2824 at \zsp\ = 0.0587 and 0.0582 \citep{Struble1999}, respectively, which are part of a filament in the Pisces-Cetus supercluster \citep{Porter2005}.

The radio core and eastern jet are also detected in NVSS, TGSS and GLEAM. The TGSS-NVSS spectral index map\footnote{https://tgssadr.strw.leidenuniv.nl/hips$\_$spidx/} \citep{deGasperin2018} shows $\alpha \sim -0.3$ in the inner few arcminutes and slighter steeper values to the east before and after the sharp bend. VLASS reveals inner jets ($<$30 arcsec in length), with the eastern jet much brighter than the western one. \\

\begin{figure*}
\centering
 \includegraphics[width=14cm]{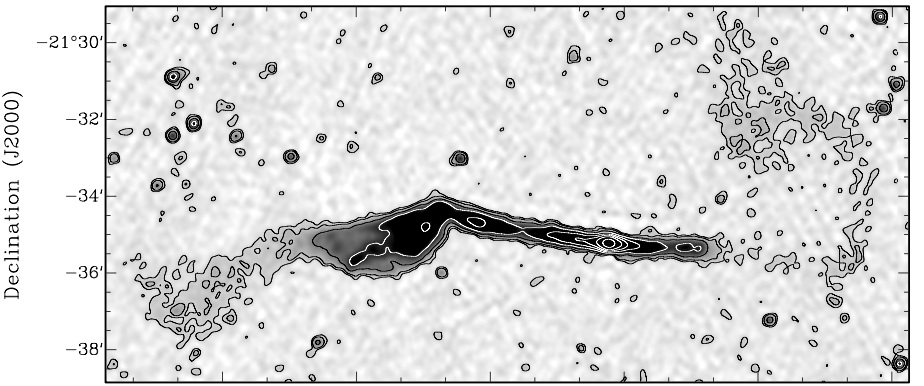} \\
 \includegraphics[width=14cm]{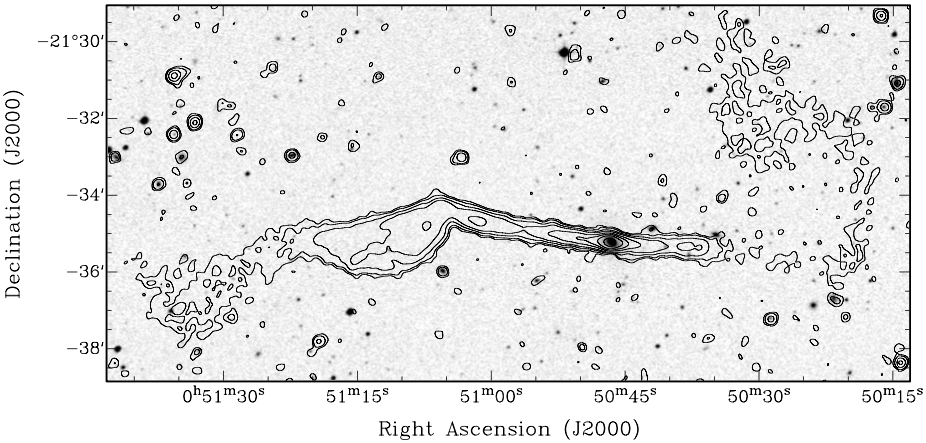} \\
\caption{ASKAP J0050--2135 (FR\,I-type GRG). --- {\bf Top}: ASKAP 944~MHz radio continuum map; the contour levels are 0.1, 0.2, 0.5 mJy\,beam$^{-1}$ (black), 1, 2.5, 5, 10 and 25.0 mJy\,beam$^{-1}$ (white). --- {\bf Bottom}: ASKAP radio contours (all black) overlaid onto a DSS2 $R$-band image. The GRG host galaxy is WISEA J005046.49--213513.6 (\zsp\ = 0.05760). The ASKAP resolution of 13 arcsec is shown in the bottom left corner.}
\label{fig:GRGJ0050-2135}
\end{figure*}

{\bf ASKAP J0050--2325} is a spectacular FR\,II-type wide-angle tail (WAT) radio galaxy consisting of a radio core, two inner jets and two diffuse, bent lobes (see Fig.~10). The optical counterpart of the radio core is clearly identified in DES images (see Fig.~10) as DES J005050.02--232509.3 (WISEA J005049.89--232511.1, 2MASX J00505000--2325097) and has a redshift of \zsp\ = 0.111367 \citep{Jones2009}. 

The whole structure, which spans around 13.5 arcmin, is rather asymmetric. Its projected linear size is $\sim$1.6 Mpc. When measured along the curved trail of radio emission the lobes are $\sim$2.4 Mpc from end to end. The western radio lobe appears to be much closer ($\sim$5 arcmin) to the core and brighter than the more extended eastern lobe ($\sim$9 arcmin). While the inner jets are linear, each extending $\sim$2.5 arcmin towards the SE and NW, and initially symmetric, the SE jet shows enhanced radio emission when it turns North before looping back to the South connecting with the SE lobe. The brightening at the end of the jet and its abrupt turn coincide with the projected location of two background galaxies (near WISEA J005059.40--232608.4) at \zph\ = 0.21 and 0.37, respectively \citep[both from][]{Zhou2021}. Because of the difference in redshift to the GRG host, we do not consider these galaxies to be physically associated with the jet.

We measure the ASKAP position of the GRG's radio core as $\alpha,\delta$(J2000) = $00^{\rm h}\,50^{\rm m}\,50.03^{\rm s}$, --23\degr\,25\arcmin\,09.32\arcsec\ (peak flux $\sim$8.8 mJy\,beam$^{-1}$). The source was previously catalogued as NVSS J005049--232509 and is also detected in VLASS as a point source ($\sim$7.5 mJy). The position of the associated WISE source, WISEA J005049.89--232511.1, is offset, likely due to confusion with a neighboring galaxy (shown in Fig.~\ref{fig:GRGJ0050-2325-DR8}) of similar redshift. \\

\begin{figure*}
\centering
 \includegraphics[width=14cm]{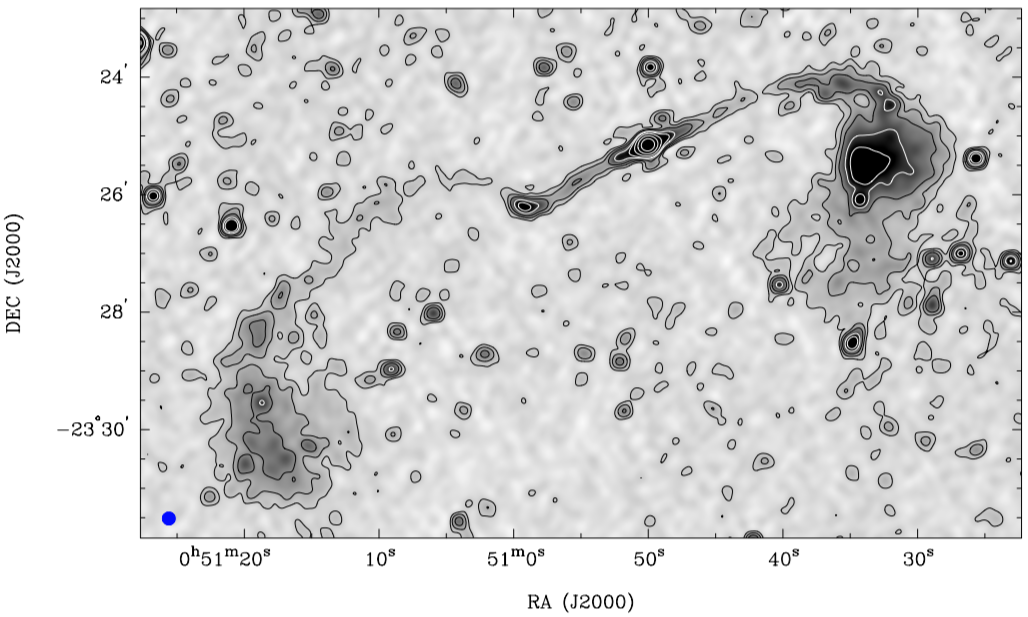} \\
 \includegraphics[width=14cm]{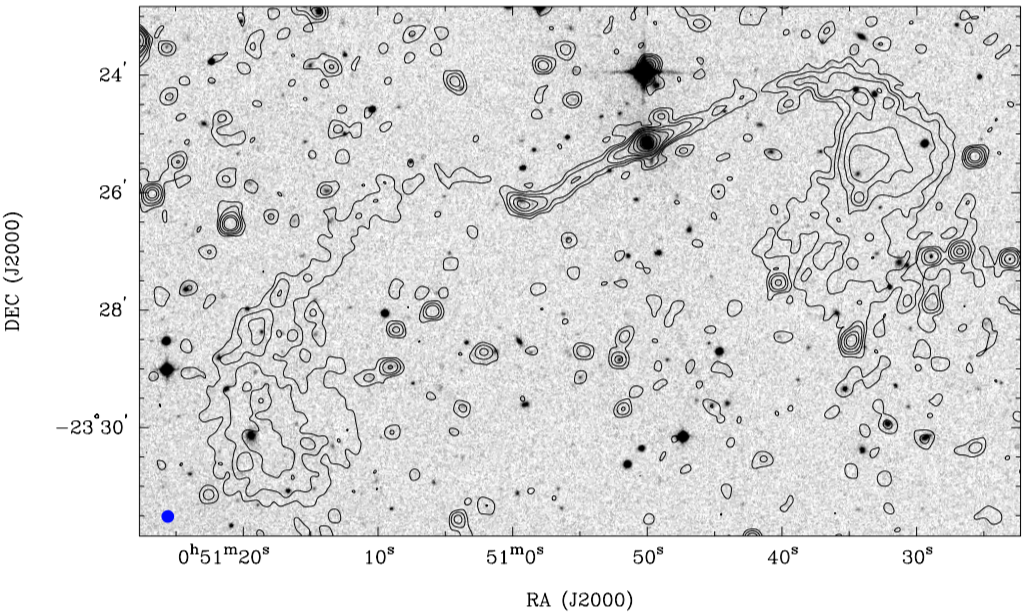} \\
\caption{ASKAP J0050--2325 (FR\,II-type GRG).  --- {\bf Top}: ASKAP 944~MHz radio continuum map; the contour levels are 0.03, 0.1, 0.2, 0.4, 0.6, 1.2, 2.5 and 5.0 mJy\,beam$^{-1}$. --- {\bf Bottom}: ASKAP radio contours overlaid onto a DSS2 $R$-band image. The GRG host galaxy is WISEA J005049.89--232511.1 (\zsp\ = 0.11137). The ASKAP resolution of 13 arcsec is shown in the bottom left corner.}
\label{fig:GRGJ0050-2325}
\end{figure*}

\begin{figure}
\centering
 \includegraphics[width=8.5cm]{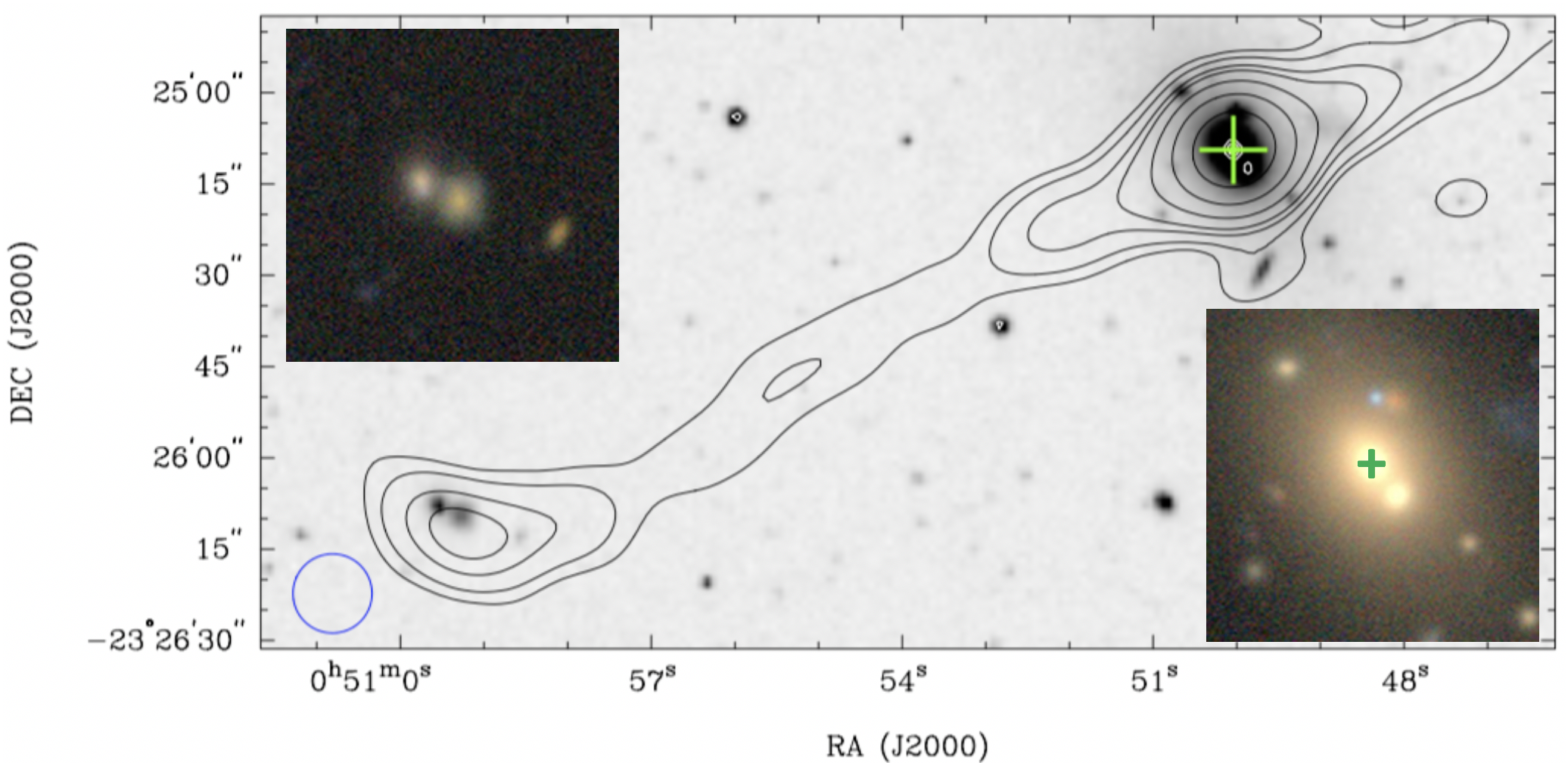}
\caption{ASKAP J0050--2325 (FR\,I-type GRG, see Fig.10). --- Zoomed ASKAP 944~MHz radio continuum map of the GRG's radio core and eastern jet. The contour levels are 0.1, 0.2, 0.4, 0.6, 1.2, 2.5 and 5.0 mJy\,beam$^{-1}$. The ASKAP resolution of 13 arcsec is shown in the bottom left corner. --- {\bf Left inset:} two background galaxies associated with WISEA J005059.40--232608.4 near the enhancement at the end of the eastern radio jet. --- {\bf Right inset:} elliptical GRG host galaxy DES J005050.02--232509.3 (\zsp\ = 0.111367); the radio core centre is marked with a green cross. The galaxy to the SW, DES J005049.87--232512.4 (\zph\ = 0.117), is an interacting companion. }
\label{fig:GRGJ0050-2325-DR8}
\end{figure}

{\bf ASKAP J0055--2231} is an FR\,I-type radio galaxy with LAS = 9.2 arcmin (see Fig.~12). The host galaxy is WISEA J005548.98--223116.9 \citep[\zsp\ = 0.11437; 6dF,][]{Jones2009}. We derive LLS = 1.14 Mpc.
This GRG has bright inner jets, fading into wider radio lobes with the western side more extended and much fainter than the eastern side. We measure flux densities of approx. 145 mJy (E lobe), 137 mJy (W lobe), and 282 mJy (total). The central radio peak (30 mJy\,beam$^{-1}$) at 944~MHz is $\sim$10 east of the host galaxy. The GRG is associated with NVSS J005549--223115 ($176.6 \pm 6.0$ mJy at 1.4~GHz) and detected in VLASS as an E--W extended source (but affected by artifacts). The TGSS-NVSS spectral index map \citep{deGasperin2018} shows much steeper values on the eastern side compared to the western side. \\

\begin{figure*}
\centering
 \includegraphics[width=14cm]{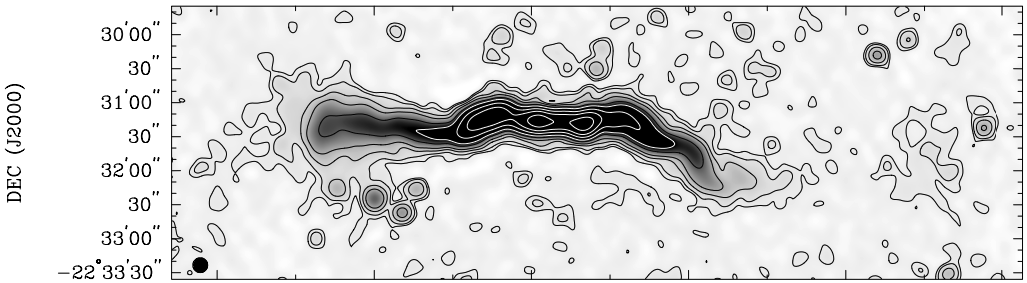} \\
 \includegraphics[width=14cm]{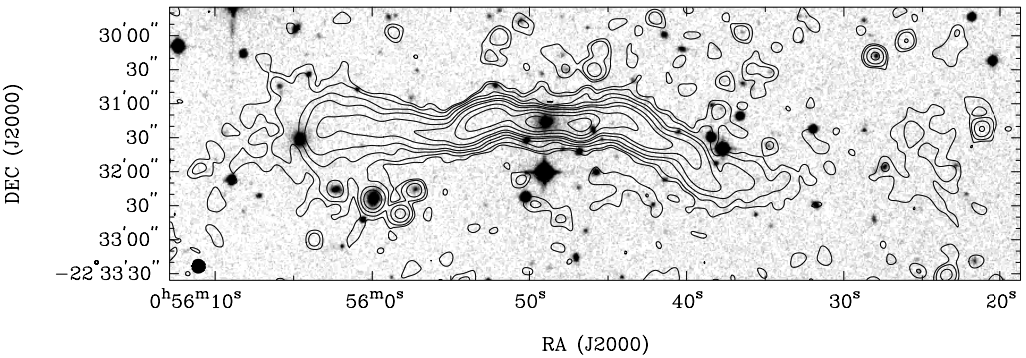} \\
\caption{ASKAP J0055--2231 (FR\,I-type GRG). --- {\bf Top}: ASKAP 944~MHz radio continuum map; the contour levels are 0.04, 0.1, 0.25, 0.5, 1.2, 2.1, 5.0, 10.0 and 20.0 mJy\,beam$^{-1}$. --- {\bf Bottom}: ASKAP radio contours overlaid onto a DSS2 $R$-band image. The host galaxy is WISEA J005548.98--223116.9 (\zsp\ = 0.11437). The ASKAP resolution of 13 arcsec is shown in the bottom left corner. }
 \label{fig:GRGJ0055-2231}
\end{figure*}

{\bf ASKAP J0057--2428} is an FR\,II-type GRG with inner radio knots (hotspots, separated by 43 arcsec) and slightly bent outer radio lobes extending 12.1 arcmin (see Fig.~13). Its host galaxy is WISEA~J005736.30--242814.9 (2MASS J00573630--2428152, DES J005736.29--242814.8) at \zph\ = $0.238 \pm 0.023$ \citep{Zhou2021}, giving LLS = 2.74 Mpc. We measure the following flux densities: 2.4 mJy (core + inner jets), 11.0 mJy (N lobe), 7.3 mJy (S lobe), and 20.6 mJy (total). The radio core is at $\alpha,\delta$(J2000) = $00^{\rm h}\,57^{\rm m}\,36.33^{\rm s}$, --24\degr\,28\arcmin\,15\arcsec\ with a peak flux of 1.5 mJy\,beam$^{-1}$ and also detected in VLASS. \\

\begin{figure}
\centering
 \includegraphics[height=7cm]{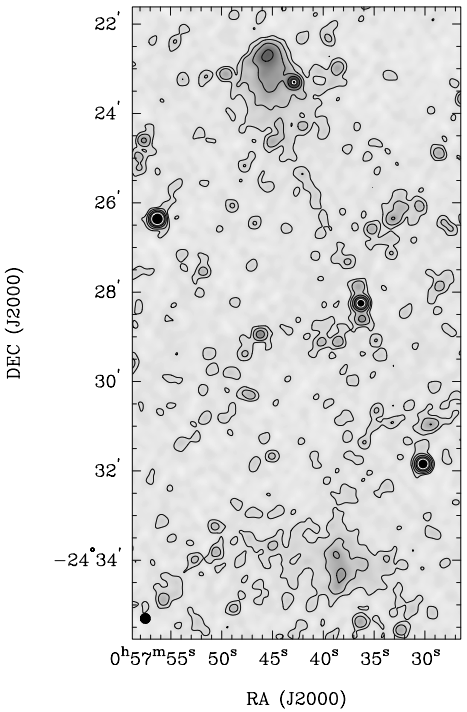}
 \includegraphics[height=7cm]{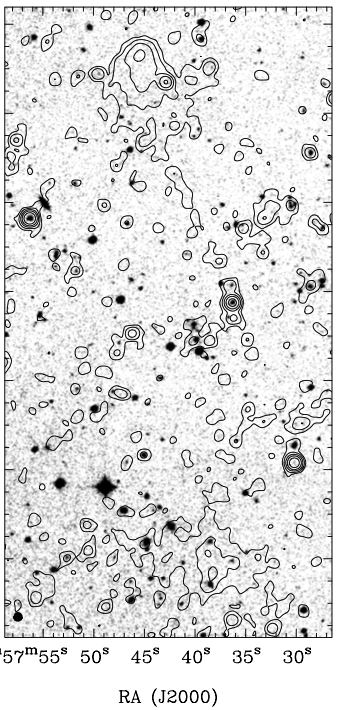} \\
\caption{ASKAP J0057--2428 (FR\,II-type GRG). --- {\bf Left}: ASKAP 944~MHz radio continuum map; the contour levels are 0.03, 0.1, 0.25, 0.5 and 1 mJy\,beam$^{-1}$. --- {\bf Right}: ASKAP contours overlaid onto a DSS2 $R$-band image. The GRG host galaxy is WISEA~J005736.30--242814.9 (\zph\ = 0.238). The ASKAP resolution of 13 arcsec is shown in the bottom left corner. }
\label{fig:GRGJ0057-2428}
\end{figure}

{\bf ASKAP J0058--2625} is a bent FR\,I/II-type GRG spanning 12 arcmin (see Fig.~14). Its host galaxy is WISEA~J005835.74--262521.3 (2MASX~J00583576--2625214; DES J005835.73--262521.2) at \zsp\ = 0.11341 \citep{Colless2001}. 
We derive LLS = 1.48 Mpc. While the inner jets are clearly detected, the outer radio lobes, particularly on the eastern side, are very faint. We measure approximate flux densities of 2.8 mJy (radio core), 2.4 mJy (E), 3.8 mJy (W), and 9.0 mJy (total). The radio core is at $\alpha,\delta$(J2000) = $00^{\rm h}\,58^{\rm m}\,35.74^{\rm s}$, --26\degr\,25\arcmin\,21.65\arcsec\ and has a peak flux of 2.2~mJy\,beam$^{-1}$. \\

\begin{figure}
\centering
 \includegraphics[width=8.5cm]{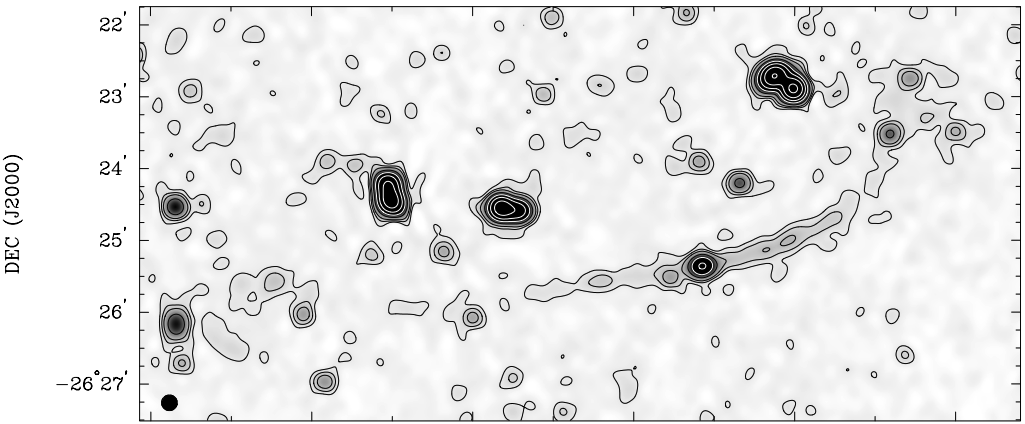} \\
 \includegraphics[width=8.5cm]{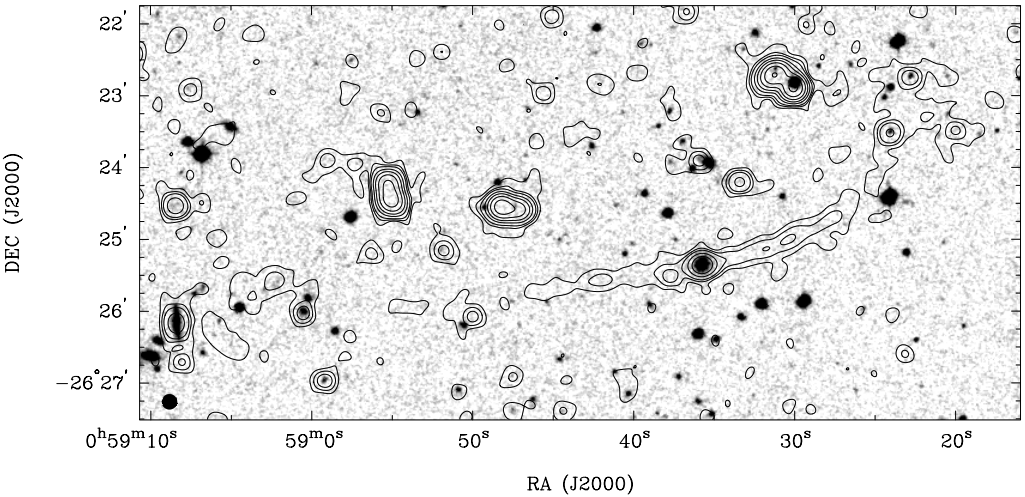} \\
\caption{ASKAP J0058--2625 (FR\,I/II-type GRG). --- {\bf Top}: ASKAP 944~MHz radio continuum map; the contour levels are 0.03, 0.1, 0.2, 0.5, 1, 2, 4, and 8 mJy\,beam$^{-1}$.  --- {\bf Bottom}: ASKAP radio contours overlaid onto a DSS2 $R$-band image. The GRG host galaxy is WISEA~J005835.74--262521.3 (\zsp\ = 0.1134). The ASKAP resolution of 13 arcsec is shown in the bottom left corner. }
 \label{fig:GRGJ0058-2625}
\end{figure}

{\bf ASKAP J0059--2352} is an FR\,II-type GRG candidate with radio lobes extending 8.0 arcmin (see Fig.~15). The association remains uncertain due to the lack of connecting jets and the presence of other bright sources near the putative lobes. Approximately midway between the latter is the potential host galaxy, WISEA J005954.72--235254.7 (DES J005954.75--235253.9), with \zph\ = $0.735 \pm 0.041$ \citep{Zhou2021}, which is our highest redshift in Table~1. The radio core is very faint compared to the bright, compact radio lobes (see Fig.~23). Based on the redshift above, we estimate LLS = 3.49 Mpc, which makes it the second largest GRG in our sample. Both lobes contain hotspots with radio emission extending towards the core and neither has optical/IR counterparts. They are also detected in NVSS and TGSS with spectral index values of around $-0.6$ and $-0.3$ for the eastern and western lobes, respectively \citep{deGasperin2018}.

Alternately, Fig.~15 may show at least two double-lobed radio galaxies, one either associated with a radio-loud quasar WISEA J010003.49--235328.5 at \zph\ = 0.14 or the galaxy WISEA J010014.11--235513.3 at \zph\ = 0.21 and the other with the early-type galaxy WISEA J005939.33--235123.8 at \zph\ = 0.26. In that case, the RGs have LAS = 3.2 arcmin (LLS = 474~kpc) and 1.0 arcmin (LLS = 240~kpc), respectively. \\

\begin{figure}
\centering
 \includegraphics[width=8.5cm]
 {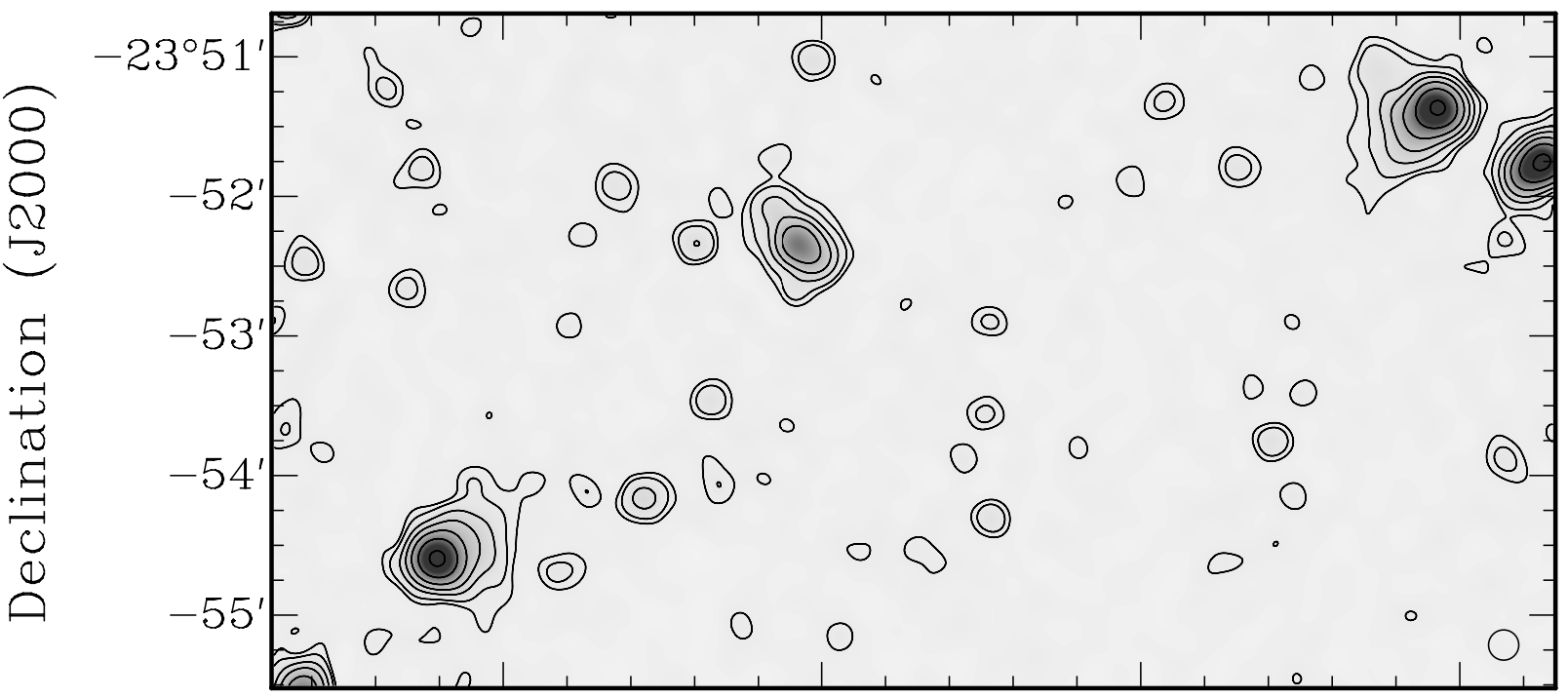}
 \includegraphics[width=8.5cm]
 {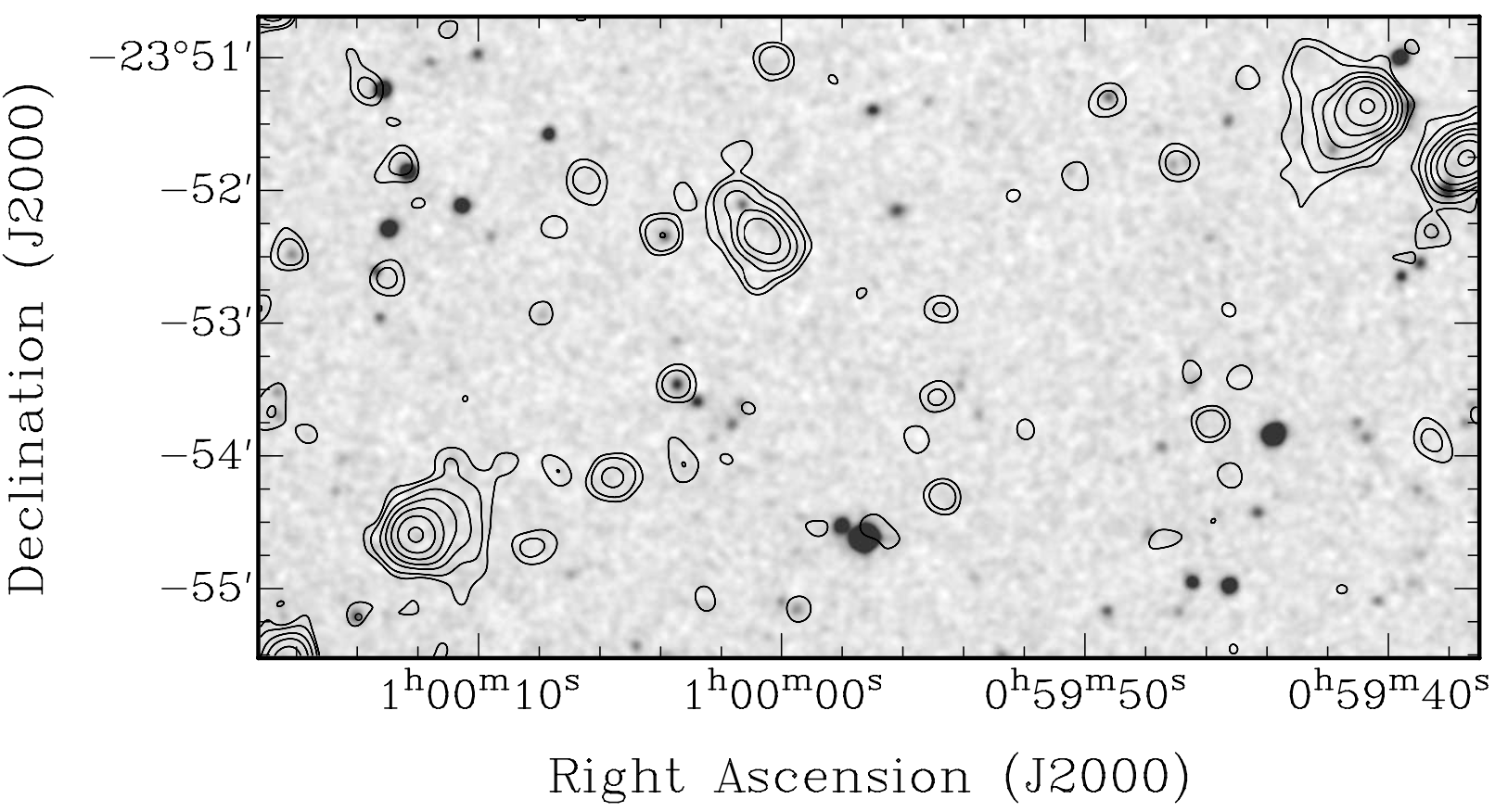}
\caption{ASKAP J0059--2352 (FR\,II-type GRG candidate). --- {\bf Top}: ASKAP 944~MHz radio continuum map; the contour levels are 0.05, 0.1, 0.25, 0.5, 1, 2, and 4 mJy\,beam$^{-1}$. The ASKAP resolution of 13 arcsec is shown in the bottom right corner. --- {\bf Bottom}: ASKAP radio contours overlaid onto a DSS2 $R$-band image. The GRG host galaxy is WISEA~J005954.72--235254.7 (\zph\ = 0.735).}
\label{fig:GRGJ0059-2352}
\end{figure}

{\bf ASKAP J0100--2125} is an FR\,I-type GRG with host galaxy WISEA J010039.00--212533.5 (\zph\ = 0.193) and LAS = 6.34 arcmin (see Fig.~16). Faint bi-polar jets connect to diffuse radio lobes, both bending by $>$90\degr\ northwards. We derive LLS = 1.21 Mpc. The bright radio core is also detected in VLASS. The ASKAP coverage for this position is currently limited to one field (SB13570; $\sim$10~h, see Fig.~1). \\

\begin{figure}
\centering
 \includegraphics[width=8.5cm]{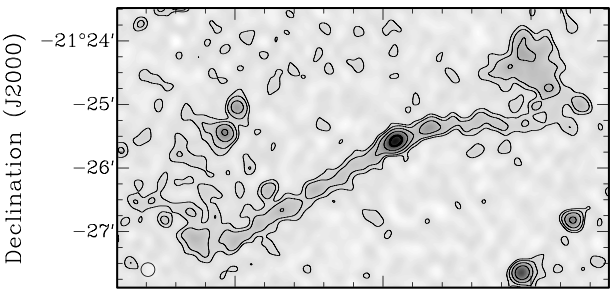}
 \includegraphics[width=8.5cm]{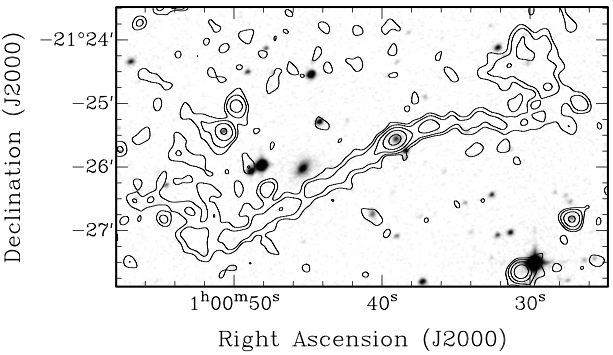}
\caption{ASKAP J0100--2125 (FR\,I-type GRG). --- {\bf Top}: ASKAP 944~MHz radio continuum map; the contour levels are 0.06, 0.12, 0.25, 0.5, and 1 mJy\,beam$^{-1}$. The ASKAP resolution of 13 arcsec is shown in the bottom left corner. --- {\bf Bottom}: ASKAP radio contours overlaid onto a DSS2 $R$-band image. The GRG host galaxy is WISEA~J001039.00--212533.5 (\zph\ = 0.193). }
\label{fig:GRGJ0100-2125}
\end{figure}

{\bf ASKAP J0102--2154} is a complex radio structure extending N--S over 6.4 arcmin (see Fig.~17). The radio emission comes from the foreground Abell~133 galaxy cluster \citep[$z = 0.0556$,][]{Struble1999}, a radio relic identified by \citet{Slee2001} just north of and associated with the cluster, and a background GRG with host 2MASX J01024529--2154137 \citep[\zsp\ = 0.2930,][]{Owen1995} and LLS = 1.68 Mpc. For a detailed multi-wavelength study of the area see \citet{Randall2010}, who expand on the radio and X-ray analysis of the northern component by \citet{Slee2001}. The source is also part of the EMU pilot study of galaxy clusters by \citet{Duchesne2024}. The ASKAP coverage for this position is currently limited to one field (SB13570; $\sim$10~h, see Fig.~1). 

The GRG's northern lobe, which is partially located behind the merging cD galaxy (ESO\,541-G013, $z = 0.057$), appears to be connected to the host galaxy by a narrow jet-like structure. Another radio jet emerges from the host to the south, twisting and connecting to the southern radio lobe which has a peculiar, not previously seen double ring / shell morphology with a cluster galaxy (WISEA J010245.32--215729.4, \citep[\zsp\ = 0.056492,][]{Smith2004} embedded. Overall, the GRG looks like a giant "Twister", somewhat resembling Hercules\,A (3C\,348), 3C\,353 and IC\,4296. The ring-like structures inside the radio lobe are possibly annular shocks (vortex rings) expanding within the jet's backflow \citep{Saxton2002,Kataoka2008,Condon2021}. 

At the position of the host, 2MASX J01024529--2154137 (WISEA J010245.22--215414.3), the DES optical images reveal a close galaxy pair, separated by only 1.3 arcsec (5.7 kpc). Furthermore, extended, banana-shaped (lensed\,?) VLASS 3~GHz emission in the core area ($PA \sim 40$ deg) lies just offset from the galaxy pair and is misaligned with the N--S structure of the large radio structure.  \\

\begin{figure*}
\centering
 \includegraphics[height=8cm]{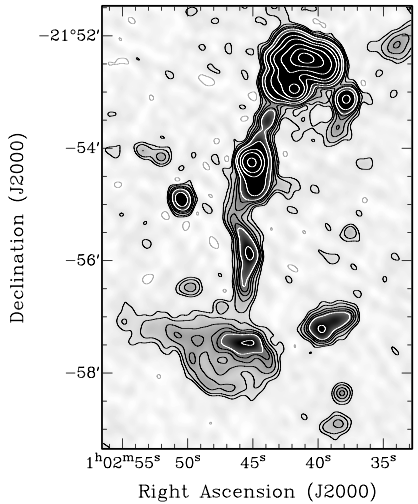} 
 \includegraphics[height=8cm]{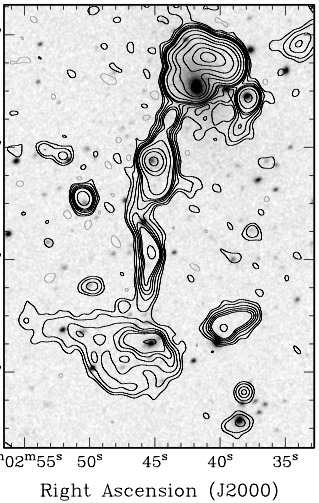}
 \includegraphics[height=8cm]{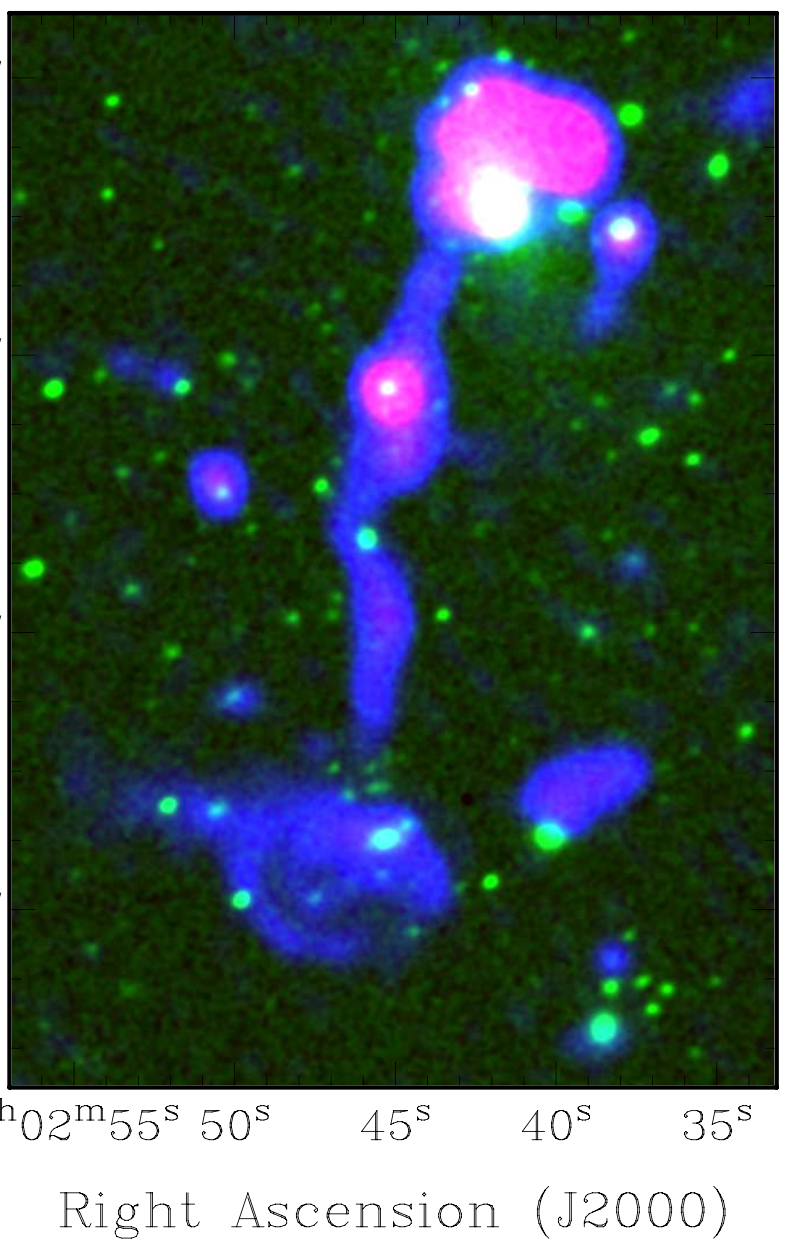} \\
\caption{ASKAP J0102--2154 (FR\,II-type GRG) and foreground Abell~133 galaxy cluster. --- {\bf Left}: ASKAP 944~MHz radio continuum map; the contour levels are 0.1, 0.2, 0.4, 0.6 and 0.8 mJy\,beam$^{-1}$ (black), 1, 2, 5, 10, 25, 50 and 80 mJy\,beam$^{-1}$ (white).  --- {\bf Right}: ASKAP radio contours (all black) overlaid onto a DSS2 $R$-band image. The GRG host is 2MASX J01024529--2154137 (\zsp\ = 0.2930). The ASKAP resolution of 13 arcsec is shown in the bottom left corner.}
 \label{fig:GRGJ0102-2154}
\end{figure*}

{\bf ASKAP J0107--2347} appears to be a re-started GRG with LAS = 13.8 arcmin (see Fig.~18). Its host galaxy is WISEA J010721.14--234734.1
(DES J010721.39--234734.0) with \zph\ = $0.312 \pm 0.024$ \citep{Zhou2021}. We derive LLS = 3.79 Mpc, which makes it the largest GRG in our sample. This GRG can also be classified as a DDRG, where  the outer lobes are relic lobes. The ASKAP coverage for this position is currently limited to one field (SB13570; $\sim$10~h, see Fig.~1). We measure the following flux densities: 5.0 mJy (core), 25.0 mJy (inner N lobe), 7.0 mJy (inner S lobe), 21.3 mJy (outer N lobe), 30.7 mJy (outer S lobe), and 89 mJy (total). The radio core is at $\alpha,\delta$(J2000) = 01:07:21.375, --23:47:34.33 with a peak flux of 4.3 mJy\,beam$^{-1}$. The radio core and inner N lobe are also detected in VLASS. The full extent of the GRG is also faintly detected in NVSS; we measure an integrated NVSS 1.4 GHz flux density of 47.5 mJy over the GRG area detected by ASKAP. The VLASS 3~GHz image shows a core of 4.2 mJy and a hotspot in the northern inner lobe, while the southern inner lobe is completely resolved out. Comparing the integrated ASKAP and NVSS fluxes we obtain a spectral index of --1.6.

A galaxy cluster at \zph $\sim$ 0.4, visible in the NE corner of Fig.~18, shows extended radio emission. For more details see Section~3.5 and Fig.~22. \\

\begin{figure*}
\centering
 \includegraphics[height=9cm]{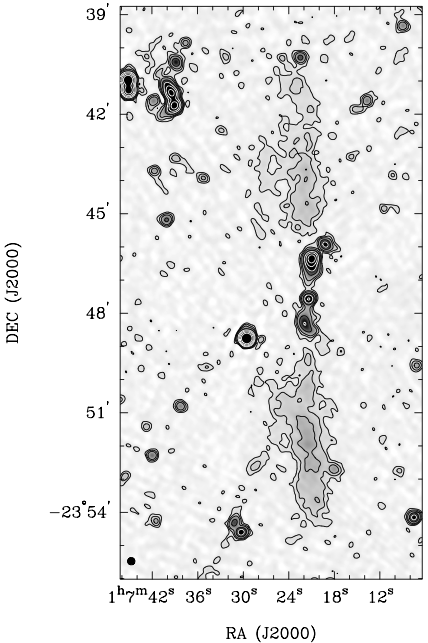}
 \includegraphics[height=9cm]{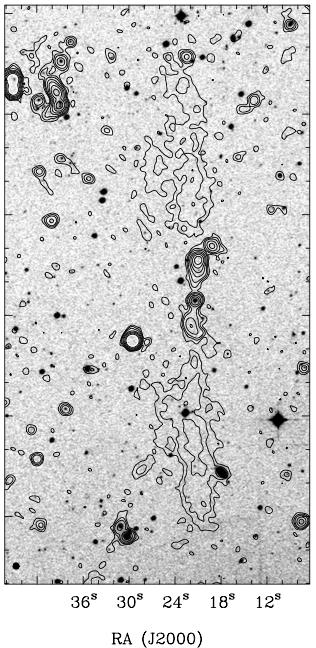}
 \hspace{0.5cm}
 \includegraphics[height=9cm]{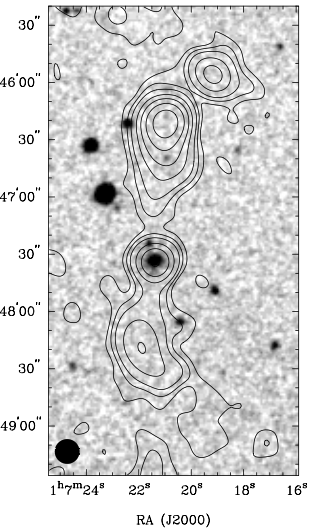}
\caption{ASKAP J0107--2347 (re-starting GRG). --- {\bf Left} ASKAP 944 MHz radio continuum map; the contours are 0.08, 0.2, 0.4, 1, 2, 4, 8, and 16 mJy\,beam$^{-1}$. --- {\bf Middle}: ASKAP radio contours overlaid onto a DSS2 $R$-band image. --- {\bf Right}: Zoom-in of the inner radio lobes. The GRG host galaxy is WISEA~J010721.14--234734 (\zph\ $\sim$ 0.312). The ASKAP resolution of 13 arcsec is shown in the bottom left corner. --- The radio-detected galaxy cluster (\zph\ $\sim$ 0.4) discovered north-east of the GRG is briefly discussed in Section~3.5.}
\label{fig:J0107-2347}
\end{figure*}

\begin{table}
\centering
\begin{tabular}{ccccccccccc}
\hline
 ASKAP & total & lobe\,1 & lobe\,2 & \multicolumn{2}{c}{core (+ inner jets)} \\
  name & \multicolumn{3}{c}{integrated flux density} & peak flux & int. flux \\
& [mJy] & [mJy] & [mJy] & [mJy/beam] & [mJy]\\
\hline
J0037--2752 & 28.2 &  6.7 (N) & 18.7 (S) & 1.52 &  2.8 \\
J0039--2541 & 68.3 &  4.8 (E) &  6.0 (W) & 19.1 & 57.5 \\
J0041--2655 & 13   &  5.3 (N) & 3.7  (S) & 
3.7 & 4.0 \\
J0044--2317 & 30.0 &  3.5 (N) & 26.0 (S) &  0.5 &  0.5 \\
J0047--2419 & 148.2 & 82.7 (N) & 56.3 (S) &  6.8 & 9.2 \\ \\
J0049--2137 & 64.1 & 40.8 (SE) & 16.7 (NW) & 5.0 & 6.6 \\
J0050--2135 & 315  & 174 (E) & 64 (W) & 45 & 77 \\
J0050--2325 & 58.2 & 16.8 (E) & 28.4 (W) & 9.2 & 13.0 \\
J0055--2231 & 282  & 145 (E) & 137 (W) & 30 & 62 \\
J0057--2428 & 20.6 & 11.0 (N) &  7.3 (S) & 1.5 & 2.4 \\ \\
J0058--2625 &  9.0 &  2.4 (E) &  3.8 (W) & 2.2 & 2.8 \\
J0059--2352 & 16.4 &  7.8 (SE) & 8.5 (NW) & $\sim$0.1 & $\sim$0.1 \\
J0100--2125 & 14.5 & 6.4 (E) & 6.0 (W) & 1.3 & 2.1 \\
J0102--2154 & 555$^{1}$ & 458$^{1}$ (N) & 44 (S) & 32 & 53 \\
J0107--2347 & 89.0 & 25.0 (N1) &  7.0 (S1) & 4.3 & 5.0 \\
            &      & 21.3 (N2) & 30.7 (S2) \\
\hline
\end{tabular}
\caption{ASKAP 944 MHz flux densities of the GRGs listed in Table~1. --- $^{1}$We subtracted 50 mJy for the approximate contribution of Abell~133.}
\label{tab:GRG-tab2}
\end{table}

\begin{figure*}
 \centering
 \includegraphics[width=3cm]{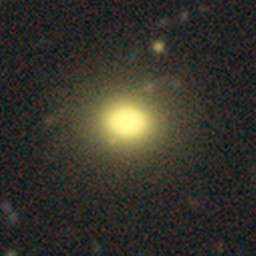}
 \includegraphics[width=3cm]{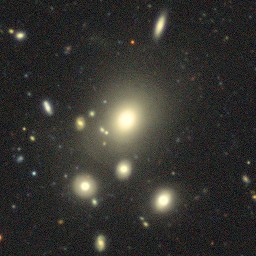}
 \includegraphics[width=3cm]{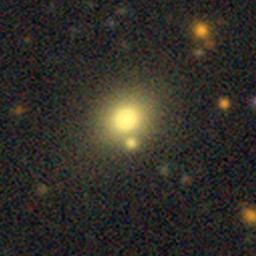}
 \includegraphics[width=3cm]
 {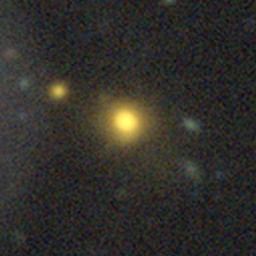}
 \includegraphics[width=3cm]{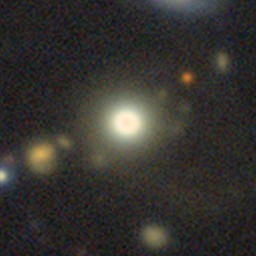} \\
 \includegraphics[width=3cm]{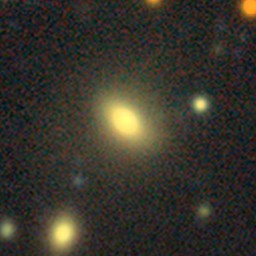}
 \includegraphics[width=3cm]{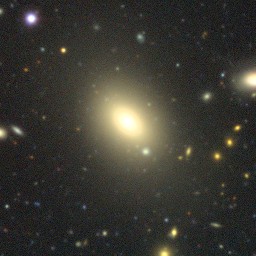}
 \includegraphics[width=3cm]{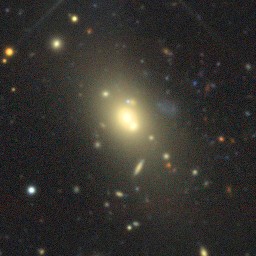} 
 \includegraphics[width=3cm]{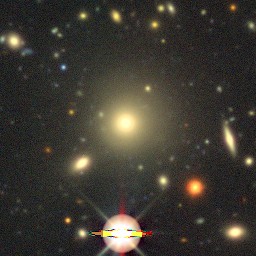}
 \includegraphics[width=3cm]{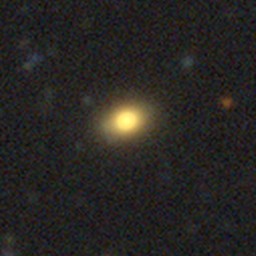} \\ 
 \includegraphics[width=3cm]{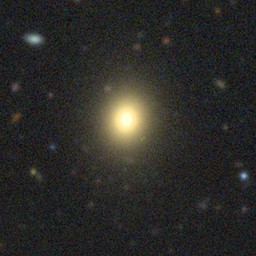} 
 \includegraphics[width=3cm]{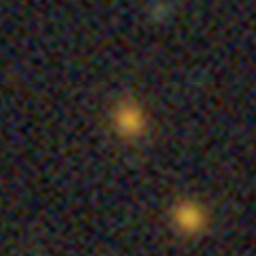} 
 \includegraphics[width=3cm]{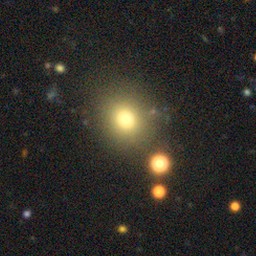}
 \includegraphics[width=3cm]{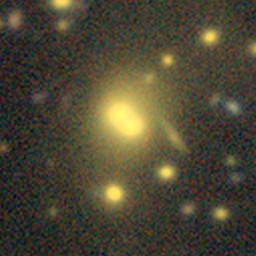}
 \includegraphics[width=3cm]{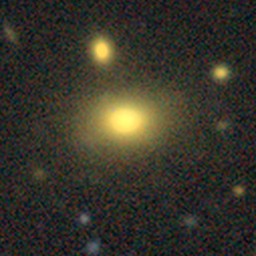} \\ 
 \vspace{0.2cm}
 \includegraphics[width=3cm]{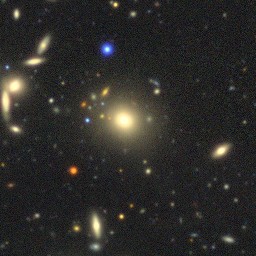}
 \includegraphics[width=3cm]{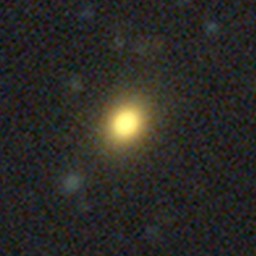}
 \includegraphics[width=3cm]{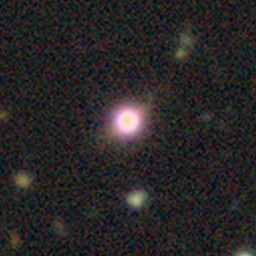}
 \includegraphics[width=3cm]{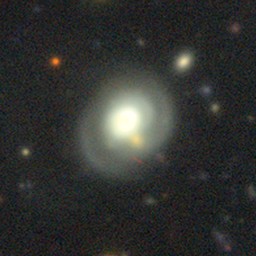} 
 \caption{Legacy Survey DES-DR10 (legacysurvey.org) optical colour images of the host galaxies of 15 GRGs (rows 1--3), and four others (last row). --- From left to right, {\bf top row:}
 WISEA J003716.97--275235.3 (\zsp\ = 0.2389),
 WISEA J003930.86--254147.8 (\zsp\ = 0.0730), 
 WISEA J004119.25--265548.3 (\zph\ = 0.232),
 WISEA J004426.72--231745.8 (\zph\ = 0.362),
 and
 WISEA J004709.94--241939.6 (\zph\ = 0.270), 
 --- {\bf second row:} 
 WISEA J004941.58--213722.1 (\zph\ = 0.233), 
 WISEA J005046.49--213513.6 (\zsp\ = 0.0576),
 WISEA J005049.89--232511.1 (\zsp\ = 0.1113).
 WISEA J005548.98--223116.9 (\zsp\ = 0.1143),
 and
 WISEA J005736.30--242814.9 (\zph\ = 0.238),
 --- {\bf third row:}
 WISEA J005835.74--262521.3 (\zsp\ = 0.1134),
 WISEA J005954.72--235254.7 (\zph\ = 0.735)
 WISEA J010039.00--212533.5 (\zph\ = 0.193)
 2MASX J01024529--21541237 (\zsp\ = 0.2930)
 and 
 WISEA J010721.41--234734.1 (\zph\ = 0.312),  
 --- {\bf bottom row:}  
 HT host galaxy, 
 WISEA J010721.41--234734.1 (\zph\ = 0.312),  
 ORC J0102--2450 host galaxy (\zsp\ = 0.27),
 WISEA J003814.72--245902.2 (\zsp\ = 0.498; QSO), 
 and the spiral DRAGN
 WISEA J004506.98--250147.0 (\zsp\ = 0.1103).}
 \label{fig:GRG-hosts}
\end{figure*}

\begin{table*}
\centering
\begin{tabular}{lccccrr}
\hline
 ASKAP & host / centre position & \multicolumn{3}{c}{WISE magnitudes} & \multicolumn{2}{c}{WISE colours} \\
 Name & $\alpha,\delta$(J2000) & W1 & W2 & W3 & W1--W2 & W2--W3 \\
 & & 3.4$\mu$m & 4.6$\mu$m & 12$\mu$m & \\
\hline
J0037--2752 & WISEA J003716.97--275235.3 & 14.05 & 13.74 & $>$12.26 & 0.31 & $<$1.48 \\
J0039--2541 & WISEA J003930.86--254147.8 & 12.32 & 12.30 & 11.57 & 0.02 & 0.73 \\
J0041--2655 & WISEA J004119.25--265548.3 & 14.81 & 14.61 & $>$12.79 & 0.20 & $<$1.82 \\
J0044--2317 & WISEA J004426.72--231745.8 & 14.73 & 14.58 & 12.65 & 0.15 & 1.93 \\
J0047--2419 & WISEA J004709.94--241939.6 & 13.05 & 12.08 & 9.74 & 0.97 & 2.34 \\ \\
J0049--2137 & WISEA J004941.58--213722.1 & 14.42 & 14.13 & $>$12.26 & 0.29 & $<$1.87 \\
J0050--2135 & WISEA J005046.49--213513.6 & 11.52 & 11.55 & 10.65 & $<$0.05 & 0.90 \\
J0050--2325 & WISEA J005049.89--232511.1 & 13.33 & 13.19 & $>$12.14 & 0.14 & $<$1.05 \\
J0055--2231 & WISEA J005548.98--223116.9 & 12.86 & 12.76 & 12.27 & 0.10 & 0.49 \\
J0057--2428 & WISEA J005736.30--242814.9 & 15.17 & 14.87 & 12.61 & 0.30 & 2.26 \\ \\
J0058--2625 & WISEA J005835.74--262521.3 & 13.33 & 13.19 & 12.21 & 0.14 & 0.98 \\
J0059--2352 & WISEA J005954.72--235254.7 & 16.77 & 16.52 & $>$12.27 & 0.25 & $<$4.25 \\
J0100--2125 & WISEA J010039.00--212533.5 & 13.74 & 13.46 & 12.04 & 0.28 & 1.42 \\
J0102--2154 & WISEA J010245.22--215414.3 & 13.78 & 13.55 & $>$12.04 & 0.23 & $<$1.51\\
J0107--2347 & WISEA J010721.41--234734.1 & 14.04 & 13.71 & 11.99 & 0.33 & 1.72 \\ 
\hline
\end{tabular}
\caption{WISE magnitudes and colours of the GRG host galaxies listed in Table~1.}
\label{tab:GRG-WISE}
\end{table*}

\subsection{A head-tail radio galaxy}
ASKAP J0055--2621 is a head-tail (HT) radio galaxy with LAS = 3.6 arcmin (see Fig.~20), and one of three HT galaxies discovered in this field. Its host galaxy is WISEA J005550.06--262155.9 \citep[\zsp\ = 0.115847,][]{Collins1995} and we derive LLS = 450 kpc. We measure a total flux density of 250 mJy of which $\sim$90 mJy are in the head area. A radio core and two inner jets (LAS = 30 arcsec, $PA \sim 170$ degr), are detected in VLASS (total flux $\sim$35 mJy, core flux $\sim$4.3 mJy), with the jets emerging perpendicular to the elliptical host galaxy. Further along, the jets merge to form a single radio tail or possibly they appear in projection along the line of sight. The HT galaxy is associated with PMN J0055--2622 and NVSS J005549--262155. It is the 2nd-brightest radio galaxy in the cluster Abell~118. \\

\begin{figure*} 
 \centering
 \includegraphics[width=14cm]{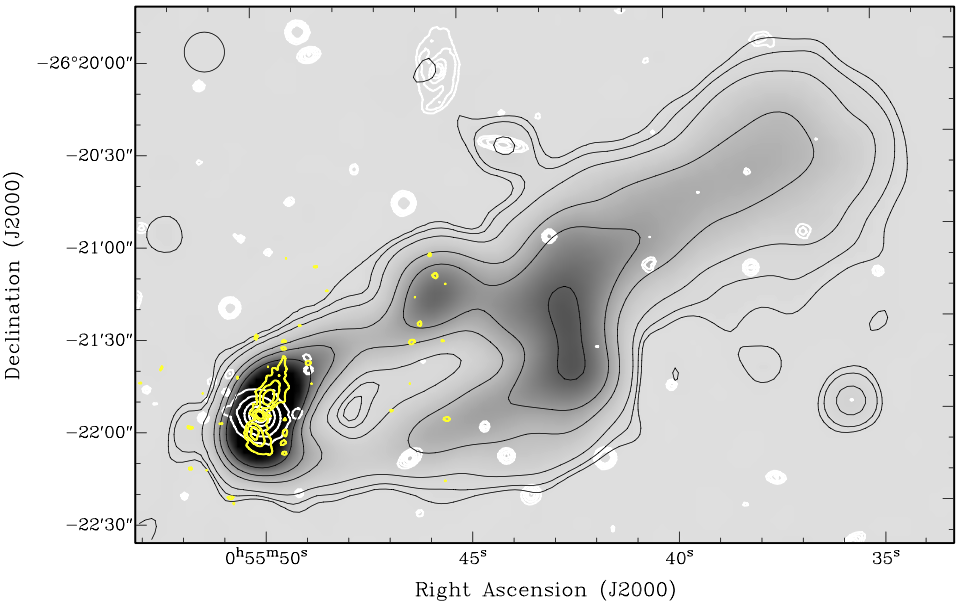}
\caption{ASKAP 944 MHz radio continuum map of the head-tail radio galaxy ASKAP~J0055--2621 with host galaxy WISEA J005550.06--262155.9 (\zsp\ = 0.115847) in Abell~118. The ASKAP contour levels are 0.1, 0.2, 0.5, 2, 4 and 8 mJy\,beam$^{-1}$. DES contours (white) and VLASS contours (yellow: 0.4, 1, 2, 4 and 8 mJy\,beam$^{-1}$) are also shown to indicate the host galaxy and inner lobes, respectively. The ASKAP resolution of 13 arcsec is shown in the top left corner.}
 \label{fig:NAT-overlay}
\end{figure*}

\subsection{A spiral DRAGN ?}
ASKAP J0045--2501 looks like a peculiar double-lobe radio galaxy with LAS = 3.5 arcmin and a bright radio core, but no prominent jets (see Fig.~21). Its host is the nearly face-on, spiral galaxy WISEA~J004506.98--250147.0 (DES~J004506.98--250146.8, LEDA~783409) with \zsp\ = 0.1103 \citep{Colless2001}, i.e. LLS = 420 kpc. Using DES optical images we measure a host galaxy diameter of $\sim$20 arcsec (40 kpc). The galaxy has a bright core/bulge ($<$5 arcsec) and a faint outer disk with spiral arms or possibly shells. Its WISE colours suggest a central, low-power active galactic nucleus (AGN) dominating the infrared emission. From the ASKAP images we measure a total flux density of $\sim$20 mJy, of which 8.7 mJy is in the radio core (peak flux = 6.8 mJy\,beam$^{-1}$). The source is also catalogued as NVSS J004507--250150 ($6.0 \pm 0.6$ mJy at 1.4~GHz). The radio core is detected in VLASS ($\sim$2 mJy), showing a N--S extension, indicating the possibility of inner jets. \\

Double-lobed Radio sources Associated with Galactic Nuclei \citep[DRAGNs;][]{Leahy1993} that have a spiral host galaxy are rare \citep[e.g.,][]{Mao2015}. Cross-matching 187\,005 SDSS spiral galaxies against extended NVSS and FIRST radio emission, \citet{Singh2015} found only four examples. A high stellar mass (and therefore large BH mass) is a defining characteristic of these spiral hosts. \citet{Bagchi2014} highlight one of the most extreme cases, a giant DDRG with a spiral host, 2MASX J23453268--0449356, and LLS = 1.6 Mpc. A prominent example of a nearby spiral with radio lobes is the Circinus Galaxy with the lobes extending $\sim$5 arcmin or $\sim$6~kpc perpendicular to the disk \citep[e.g.,][]{Elmouttie1998, Wilson2011}. Circinus is a rather isolated star-forming galaxy with a central AGN, whose radio lobes are comparable in size to its stellar disk, resembling somewhat the "radio bubbles" in the Seyfert 1.5 galaxy Mrk\,6 \citep{Kharb2006}. \\

While the location of the spiral DRAGN, ASKAP J0045--2501, suggests it could be a host galaxy candidate for GW190814, the galaxy's luminosity distance of 523~Mpc puts it beyond the event's estimated distance range of 196 -- 282~Mpc \citep{Abbott2020}. 

\begin{figure*}
\centering
 \includegraphics[height=6.5cm]{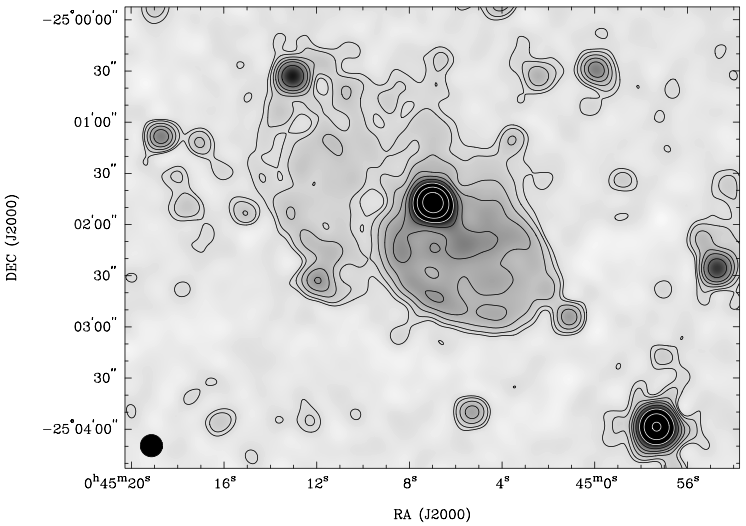}
 \includegraphics[height=6.5cm]{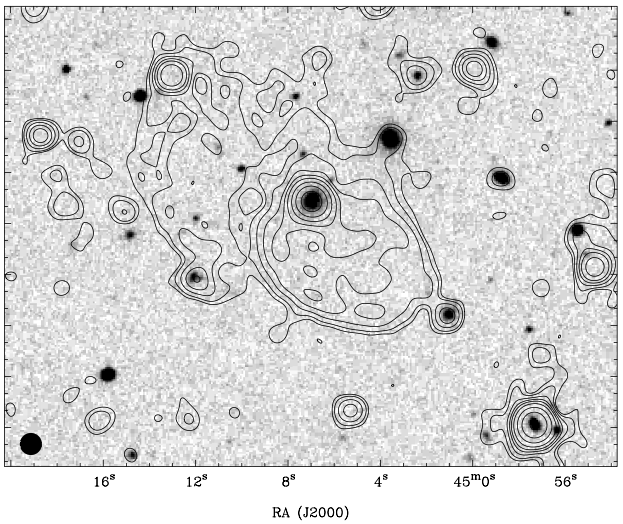}
\caption{ASKAP J0045--2501. --- {\bf Left:} ASKAP 944 MHz radio continuum map; the contour levels are 0.03, 0.06, 0.12, 0.2, 0.3, 0.5, 1, 2, 4, and 8 mJy\,beam$^{-1}$. --- {\bf Right:} ASKAP radio contours overlaid onto a DSS2 $R$-band image. The bright central host is the spiral galaxy WISEA~J004506.98--250147.0 (\zsp\ = 0.1103, see Fig.~19), which makes this a rare spiral DRAGN. The ASKAP resolution of 13 arcsec is shown in the bottom left corner.}
\label{fig:J0045-2501}
\end{figure*}

\subsection{ORC J0102--2450}

The first odd radio circle (ORC J2103--6200) was discovered in ASKAP 944~MHz data from the EMU pilot survey, followed by ORC~1555+2726 in GMRT 325~MHz data, both reported in \citet{Norris2021-ORC}. A third single odd radio circle (ORC~J0102--2450) was found by  \citet{Koribalski2021} in the deep ASKAP field studied here. 

The central radio source of ORC~J0102--2450 is associated with the bright elliptical galaxy 2MASS J01022435--2450396, which has a photometric redshift of \zph\ $\approx$ 0.27 \citep{Bilicki2016, Zou2019}, recently confirmed by \citet{Rupke2024}. ORC~J0102--2450 has a diameter of $\sim$70~arcsec corresponding to 300~kpc at the host galaxy redshift, similar in size to the first two ORCs. The discovery of ORC~J0102--2450, only the third single ORC, established the importance of their massive central galaxies ($M_{\star} \gtrsim 10^{11}$\Msun) for their formation. A possible scenario involving outwards moving merger shocks, which occur during the formation of the massive elliptical host, is presented by \citet{Dolag2023} and further explored in   \citet{Koribalski2024-ORC,Koribalski2024-Physalis}.

Searches for ORCs in radio images are very much encouraged as increasing their numbers is essential for establishing their properties and understanding their formation mechanisms. Several groups \citep[e.g.,][]{Gupta2022, Segal2023, Lochner2023, Stuardi2024} use machine learning algorithms to search for complex / anomalous radio sources, including ORCs. Nevertheless, by eye searches are currently yielding the majority of ORCs and ORC candidates, e.g. , ORC~J1027--4422 \citep{Koribalski2024-ORC}, Physalis \citep{Koribalski2024-Physalis}, and ORC~J0219--0505 \citep{Norris2025}. \\

The closest GRGs to ORC J0102--2450 are the 12.1 arcmin long FR\,II galaxy, ASKAP J0057--2428 (\zph\ = 0.238), which spans $\sim$3~Mpc, and the 15~arcmin long re-started DDRG, ASKAP J0107--2347 (\zph\ = 0.312), with a linear size of nearly 4~Mpc.

\subsection{Galaxy clusters}

There are 18 Abell clusters with known redshifts in the ASKAP Sculptor field, which can be grouped into three redshift ranges: $z \sim 0.06$ (7), $z \sim 0.11$ (4), and $z > 0.16$ (7), corresponding to filaments in the Pisces-Cetus Supercluster \citep{Porter2005}, the "Farther Sculptor Wall" \citep{Zappacosta2010}, and beyond. These large-scale structures are located well behind the better known Sculptor Wall ($z \sim 0.03$) and loose Sculptor galaxy group ($D$ = 2 -- 5~Mpc.), which includes the starburst galaxy NGC~253 (prominent in Figs.~1 \& 24). 

We detect notable diffuse radio emission in the galaxy clusters Abell~114 ($z \sim 0.06$), Abell~118 (see Section 3.2), Abell~122 ($z \sim 0.11$), Abell~133 ($z \sim 0.06$, see Section~3.1 where it is discussed together with the background GRG ASKAP J0102--2154), Abell~140 ($z \sim 0.16$; two radio relics), Abell~141 \citep[$z \sim 0.23$; radio halo][]{Duchesne2024}, and Abell~2800 ($z \sim 0.06$, see Table~5). Furthermore, we detect diffuse radio emission around the SPT-CL J0049--2440 cluster \citep[$z \sim 0.53$,][]{Hilton2021}. There is also diffuse radio emission, extended $\sim$1.2 arcmin N--S, near $\alpha,\delta$(J2000) = $01^{\rm h}\,01^{\rm m}\,43.5^{\rm s}$, --20\degr\,40\,arcmin\,16\arcsec, but no optical/IR host candidate.

During our study of the giant radio galaxy ASKAP J0107--2347 (see Fig.~18), we discovered a background galaxy cluster at $\alpha,\delta$(J2000) = $01^{\rm h}\,07^{\rm m}\,40^{\rm s}$, --23\degr\,41\arcmin\,35\arcsec\ with extensive radio emission from individual galaxies (\zph\ $\sim$ 0.4) as well as connecting filaments. An overlay of the ASKAP contours on a DES $g$-band image is shown in Fig.~22. The radio emission is also detected in NVSS \citep[see, e.g.,][]{Zanicelli2001} and GLEAM.

\begin{figure}
\centering
\includegraphics[width=8cm]{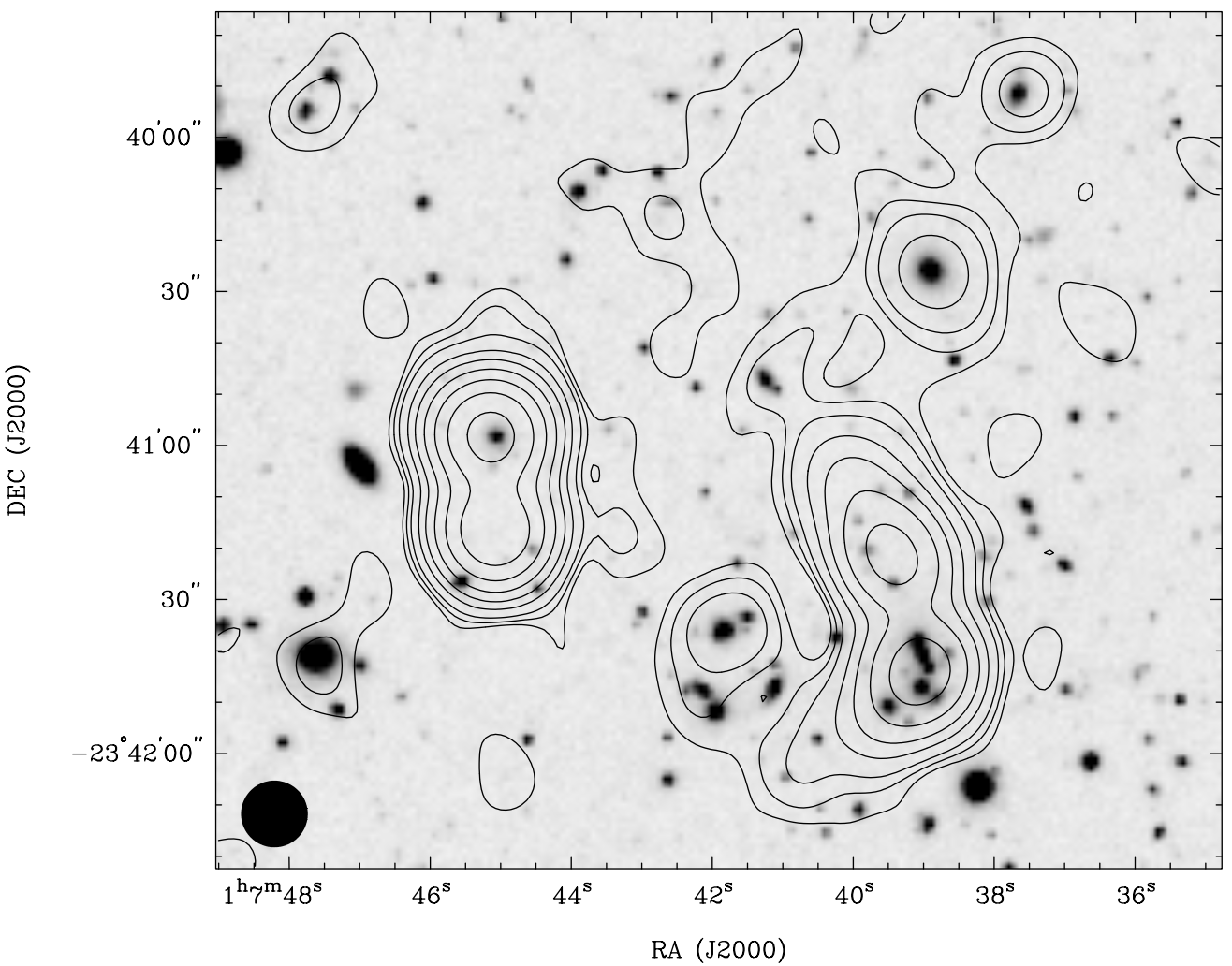}
\caption{ASKAP 944 MHz radio continuum contours overlaid onto a DES $g$-band image, revealing a galaxy cluster ($z \sim 0.4$); see also Fig.~13. The ASKAP resolution of 13 arcsec is shown in the bottom left corner.}
\label{fig:cluster}
\end{figure}

\begin{figure*}
\centering
 \includegraphics[width=18cm]{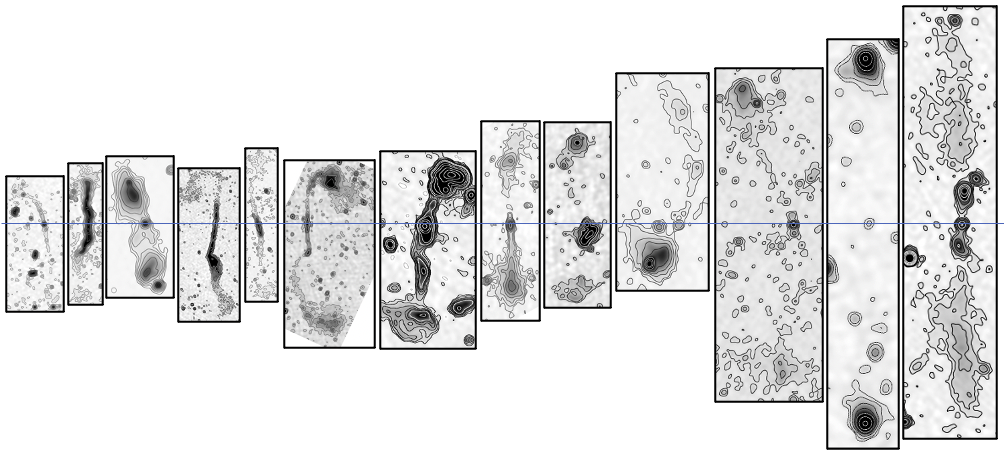}
\caption{Giant radio galaxies and candidates from Table~1 sorted by their projected linear sizes from 1.1 Mpc (left) to 3.8 Mpc (right). For display purposes the ASKAP radio continuum images are cropped and rotated, and the GRG cores are aligned along the overlaid horizontal line. }
    \label{fig:grgs-by-size}
\end{figure*}


\section{Discussion}

The high-sensitivity ASKAP 944~MHz radio continuum image of the LIGO--NGC~253 field (resolution 13\arcsec, rms noise sensitivity $\gtrsim$10~$\mu$Jy\,beam$^{-1}$) revealed 15 GRGs with LAS $>$ 5 arcmin. Small ASKAP images mark their locations within the $\sim$40~deg$^2$ area displayed in Fig.~1. In Fig.~\ref{fig:grgs-by-size} these images are shown side-by-side, in order of their LLS, which range from 1.1 to 3.8~Mpc. This highlights the wide spectrum of morphologies, including high surface brightness components such as the core and inner jets / lobes as well as very low surface brightness components such as outer remnant lobes. While some GRGs appear symmetric in shape and brightness, most show significant asymmetries and bending. The inner jets typically forge a straight path through the surrounding medium. In contrast, the volume of the fading remnant lobes expands as they become buoyant. Their 3D shapes reflect the variations in density of the surrounding medium as well as its turbulence, shocks and other motions \citep[e.g.,][]{Eilek1984,Eilek2002,Oei2022,Koribalski2024-Corkscrew}. The duty cycle can be directly constrained for ‘double–double radio galaxies’ \citep{Schoenmakers2000} and some remnant radio galaxies (Turner 2018), or estimated on a population level from the radio-loud fraction (e.g. Best et al. 2005; Shimwell et al. 2019).

In Fig.~24 we highlight the locations of the 15 GRGs with respect to the known Abell clusters. Notably, ASKAP J0050--2135 lies between A114 and A2800, which are part of a filament in the Pisces-Cetus supercluster, while ASKAP J0058--2625 lies in the vicinity of A122 and A118. In future, a deep X-ray study of this field would be of interest to further investigate the large-scale environment of GRGs. We list the radio sources with ROSAT X-ray detections in Table~\ref{tab:rosat}, several of which were also recorded by \citet{Mahony2010}. \\

Our GRG discoveries in the southern Sculptor field add less than 0.1\% to the rapidly growing GRG catalogues. Notably, the vast majority of catalogued GRGs reside in the northern hemisphere, found in LOTSS 144~MHz images \citep{Mostert2024}, showing the strong need for GRG catalogues in the southern hemisphere. While our sample is small, it highlights a wide range of radio morphologies, sizes, and surface brightness structures. On-going ASKAP radio continuum surveys such as EMU will deliver vast GRG catalogues in the southern hemisphere. Based on the current density, we estimate $\sim$20k GRGs with LSS $>$ 0.7~Mpc in the EMU southern sky survey (see Table~\ref{tab:surveys}).

\subsection{ASKAP J0107--2347}
This is with LLS = 3.8~Mpc the largest GRG in our sample, consisting of a pair of double radio sources with a common centre. It is characterized by the presence of a 550-kpc large edge-brightened, double radio source which is situated within, and well aligned with larger (3.8~Mpc) radio lobes, with both sources originating from the same host galaxy. The radio spectrum of the outer remnant lobes is steeper than that of the more recent, inner lobes. The `double-double' nature of this giant radio galaxy, points at a short interruption (a few Myr) of the jet activity. The outer lobes haven't yet faded away and the inner lobes are already well developed. The inner lobes (N1 \& S1, together 32~mJy) show only a small misalignment ($\sim$10 degr) with the outer lobes (N2 \& S2, together 52~mJy), see Table~2. 

\begin{table*}
\centering
\begin{tabular}{ccccc}
\hline
 Name & redshift & type & 1RXS \\
\hline
 NVSS J003736--230223 & 0.304 & G & J003736.3--230228 \\ 
 NVSS J003814--245902 & 0.498 & QSO & J003815.8--245858\\
 NVSS J004016--271912 & 0.172 & G & J004016.5--271913 \\
 NVSS J004539--224354 & 1.537 & QSO & J004540.4--224402 \\
 NGC~253 & 0.001 & starburst galaxy & J004733.3--251722 \\
 NVSS J004856--223304 & ? & QSO? & J004855.3--223225 \\
 NVSS J005446--245529 & 0.610 & QSO, BLLac & J005447.2--245532 \\
 TON S180 & 0.062 & Sy\,1 spiral galaxy & J005720.4--222300 \\
 Abell~122 & 0.113 & cluster & RX J0057.4--2616 \\
 ESO\,541-G013 & 0.057 & Sy\,1 galaxy in Abell~133 & J010242.8--215250 \\
 NVSS J010250--200155 & 0.370 & QSO? & J010251.3--200154 \\
 NVSS J010256--264637 & 1.606 & QSO & J010255.8--264637 \\
 Abell~141 & 0.229 & cluster & RXC J0105.5--2439 \\
 NVSS J010645--235805 & ? & QSO? & J010646.1--235800 \\
 CS Cet & Galactic & variable star & J010649.2--225122 \\
\hline
\end{tabular}
\caption{Radio sources with ROSAT X-ray detections}
\label{tab:rosat}
\end{table*}

\begin{figure*}
\centering
    \includegraphics[width=0.9\linewidth]{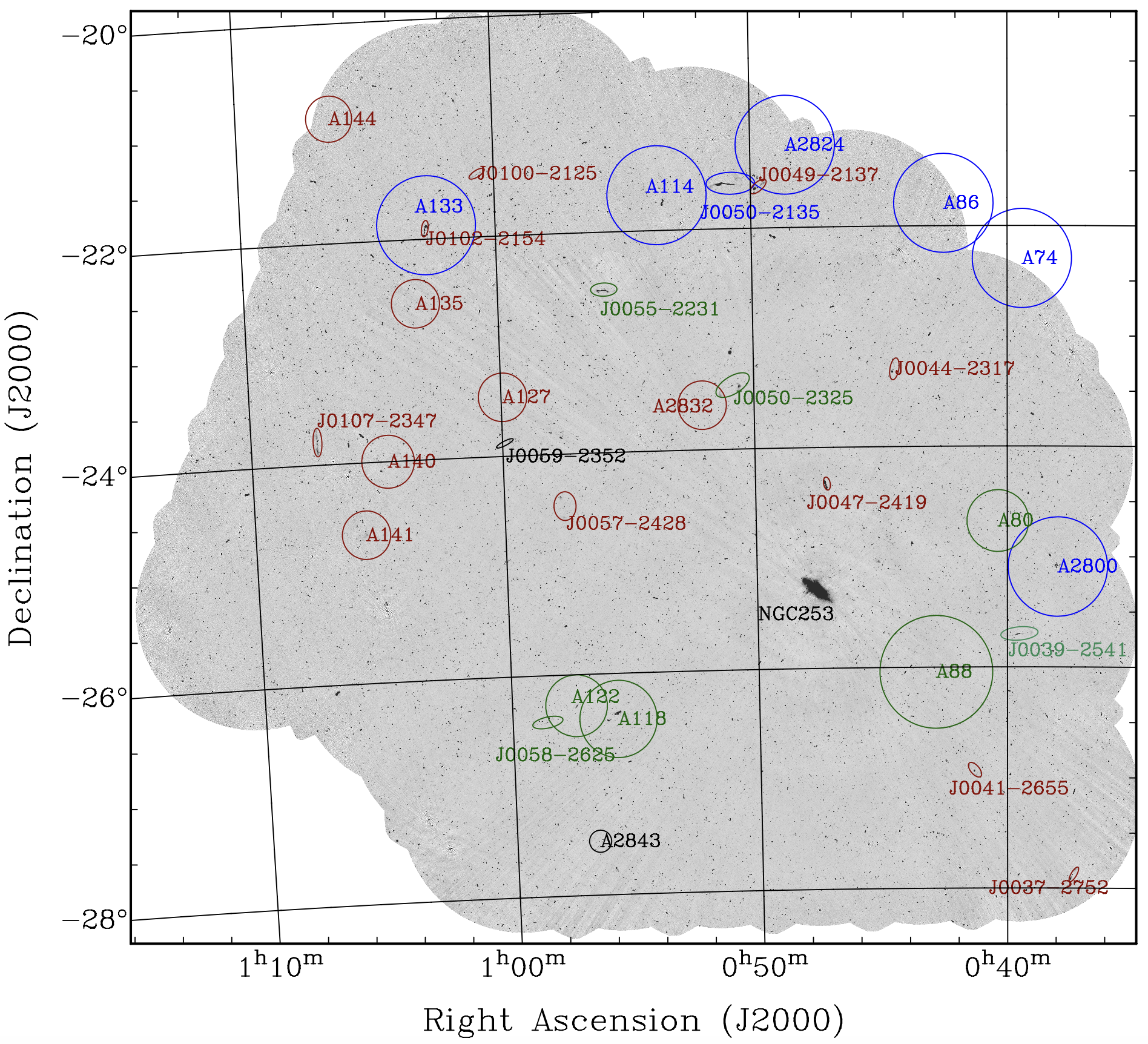}
\caption{Similar to Fig.~1, here with labels added for known Abell clusters in the field and the 15 GRGs from Table~1. }
\label{fig:field+labels}
\end{figure*}

\section{Summary and Outlook}

\begin{figure*}
\centering
    \includegraphics[width=13cm]{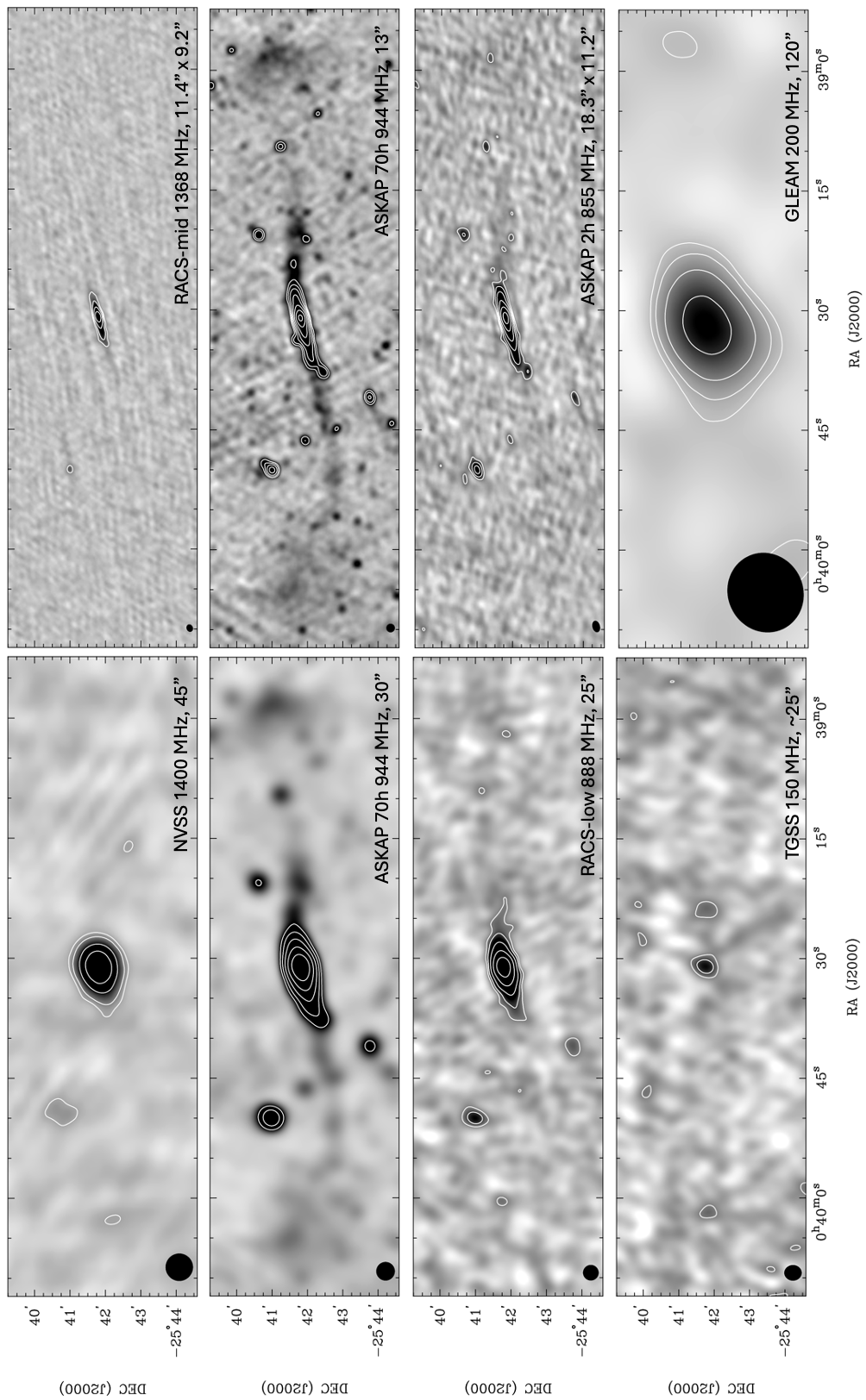}
\caption{Interferometric radio continuum images of the FR\,I-type giant radio galaxy ASKAP J0039--2541 (see Fig.~3; LAS = 15.5 arcmin and LLS = 1.29~Mpc) at frequencies from 150 to 1400 MHz for a range of angular resolutions and sensitivities. The synthesized beam is shown in the bottom left and the telescope / survey, frequency and beam size are given in the bottom right of each panel. The outer (remnant) radio lobes are only detected in the deep ASKAP 944~MHz images presented here (second row), and the emission can be traced all the way to the core. The inner region, consisting of the radio core and active jets, is detected in NVSS, RACS and GLEAM, but only resolved in RACS, while the core is also detected in TGSS.}
\label{fig:comparison-collage}
\end{figure*}

Our search for GRGs with large angular sizes (LAS $>$ 5 arcmin) in the $\sim$40 deg$^2$ ASKAP Sculptor field resulted in 15 sources with a wide range of morphologies (4$\times$ FR\,I, 8$\times$ FR\,II, one HyMoRS, and two FR\,I/II relic). For a summary of their properties see Tables~1--3. Among these are two candidate GRGs which require confirmation. The GRG angular sizes range up to 18 arcmin, with host galaxy redshifts between $\sim$0.06 and 0.36 (0.74), and their projected linear sizes range from 1.1 to 3.8~Mpc. ASKAP images of these 15 GRGs are shown in Figs.~1--18, and a side-by-side comparison is shown in Fig.~\ref{fig:grgs-by-size}. Notably, the percentage of FR\,I and mixed type RGs compared to FR\,II-type RGs is larger in our LAS $>$ 5 arcmin sample than in the full catalogue which is dominated by FR\,II-type RGs (70\%). For one of our GRGs, ASKAP J0039--2541, we present an 8-panel comparison of interferometric radio continuum images in Fig.~\ref{fig:comparison-collage}, highlighting the need for high low-surface brightness sensitivity and high angular resolution to study GRGs and other extended radio sources (e.g., radio relics). \\

The largest GRG in our sample (ASKAP J0107--2347), with LAS = 13.8 arcmin and LLS = 3.8~Mpc (host galaxy redshift \zph\ = 0.312), is an FR\,II-type DDRG. Its newly-formed inner lobes, which already span 2.2 arcmin ($\sim$600 kpc), are bright and compact, while the outer relic lobes are elongated and of very low surface brightness. While the radio core and the hot spots of the inner lobes are detected in RACS-low, RACS-mid, and NVSS, the outer relic lobes are only marginally detected. The combination of ASKAP's high resolution and good surface brightness sensitivity is likely to reveal many more relic lobes as well as other LSB radio structures. Relic lobes are of particular importance, as they give insights into the timescale of SMBH activity and are often found in the outskirts of re-started RGs. This suggests that the measured angular sizes of known radio galaxies will grow as previously undetected relic lobes are discovered both for FR\,I-type and restarting radio galaxies, increasing the numbers for these morphology types. For FR\,II-type galaxies, the physical connection between the radio core and double lobes may in many cases only be established at higher sensitivity. \\

We also find one spiral DRAGN and several examples of diffuse radio emission (halos, relics) in galaxy clusters. The discovery of ORC J0102--2450 was presented in \citet{Koribalski2021}. We note that the location and redshift of the galaxy merger system ESO\,474-G026 make it a possible host of GW190814, but no radio variability was noted in the ASKAP data. \\

In total we catalogued 232 radio galaxies whose properties are summarised in the Appendix (see Table~\ref{tab:appendix}). Of these, 77 are larger than 0.7~Mpc and 35 larger than 1~Mpc. Interpolating these numbers to the whole southern sky suggests at least 20\,000 (40\,000) radio galaxies larger than 0.7 (1) Mpc. While EMU is not as deep as the ASKAP Sculptor field presented here (80h integration vs 10h for EMU), a large fraction of these will be detectable. A combination of dedicated machine learning tools, visual inspection, and optical/infrared cross identification will be required to catalogue and verify GRG candidates. Furthermore, cross matching with X-ray cluster catalogues will be useful to explore the GRG environments. \\

While GRGs appear to be rare, their number density is likely much larger than currently estimated. \citep[][LOTSS DR2]{Mostert2024} catalogued 11485 (4979) RGs with estimated linear size larger than 0.7 (1) Mpc, resulting in densities of 2.0 (0.9) per square degree (similar to our ASKAP Sculptor field). While only 408 (311) are located in the southern hemisphere, this number will grow rapidly as several ASKAP radio continuum surveys cover the southern sky at frequencies between 0.7 and 1.6~GHz.

The better the LSB sensitivity of large-area interferometric radio surveys (see Table~\ref{tab:surveys}), the more fading / aging lobes will be detected, typically found beyond a new set of jets or lobes. This means that a fraction of catalogued "normal-sized" radio galaxies will actually be giants upon detection of their remnant lobes. Given the life cycle of RGs (active, dying / remnant, restarting or re-energised by collision) fading remnant lobes \citep[radiative and adiabatic losses,][]{Godfrey2017} should exist beyond the active jets / hot spots of all RGs. For several reasons, these old lobes are hard to detect as their spectral indices become steeper, their emission fainter, their morphology more amorphous and their volume-filling factor larger. Compelling examples of dying radio galaxies in group environments are NGC~1534 \citep{Hurley2015, DJH2019} and SGRS J0515--8100 \citep{Sub2006}. We look forward to many more discoveries of exciting radio sources, including GRGs and ORCs, and radio source statistics from the on-going ASKAP surveys.

\begin{table*}[]
\centering
\begin{tabular}{cccccccccc}
\hline
 & & & & & & GRGs $>$ 0.7 Mpc \\
 survey & telescope & frequency & beam & median rms & tint & total (rate) REF & total (rate) REF \\
 & & [MHz] & [arcsec] & [mJy\,beam$^{-1}$] & [min.] & [--, deg$^{-2}$, --] & [--, deg$^{-2}$, --] \\
\hline
\hline 
 TGSS ($\delta > -53\deg$) & GMRT & 150 &  $\gtrsim$25 & 3.5 & 15 & 673 (0.02) B24 & 0.6M (17) I17 \\
 GLEAM ($\delta < +30\deg$) & MWA & 200 & 120 & $\sim$15 & & & 0.3M (13) HW17 \\
 SUMSS ($\delta < -30\deg$) & Molonglo & 843 & $\gtrsim$45 & $\sim$1 & 1 & & 0.2M (20) M03 \\
 FLASH ($\delta \lesssim 0\deg$) & ASKAP & 856 & 15 & 0.090 & 120 & & \\
 RACS-low ($\delta < +41\deg$) & ASKAP &  888 & $18.4 \times 11.6$ & 0.266 & 15 & $\sim$13k ($\sim$0.5) A21 & 2.3M (67) D25 \\ 
 RACS-low2 ($\delta < +48\deg$) & ASKAP &  888 & $\sim$15 & $\sim$0.2 & 15 & & \\
 RACS-low3 ($\delta < +48\deg$) & ASKAP &  944 & $13.4 \times 11.0$ & 0.205 & 15 & & \\
 EMU ($\delta \lesssim 0\deg$) & ASKAP &  944 & 15 & 0.030 & 600 & $\sim$20k (1.0) estimate & $\sim$15M (750) estimate \\
 {\bf N253 field ($\sim$40 deg$^2$)} & ASKAP &  944 &  13 & $\gtrsim$0.013 & $\leq$4800 & {\bf 77 (1.9) here} & n/a & \\
 RACS-mid ($\delta < +49\deg$)  & ASKAP & 1368 & $10.1 \times 8.1$ & 0.198 & 15 & & 3.1M (86) D25 \\ 
 Wallaby ($\delta \lesssim 0\deg$) & ASKAP & 1368 & $\sim$8 & check & 960 \\
 RACS-high ($\delta < +48\deg$) & ASKAP & 1656 & $8.6 \times 8.3$ & 0.209 & 15 & & 2.7M (72) D25 \\
 NVSS ($\delta > -40\deg$) & VLA & 1400 & 45 & $\sim$0.45 & 0.4 & 10 (0.03) D17 & 1.8M (52) C98 \\
 VLASS ($\delta > -40\deg$) & VLA & 3000 & 2.5 & $\sim$0.1 & & 31 ($<$0.01) G23 & 1.9M (56) G21 \\ 
\hline
\end{tabular}
\caption{Selection of large interferometric radio continuum surveys. ASKAP surveys like EMU, Wallaby, and FLASH, are ongoing and will likely be extended further north. -- References: A21: \citet[][extrapolated GRG count]{Andernach2021}, \citet{Bhukta2024}, \citet{Condon1998}, D17 \citet{Dabhade2017}, D25: \citet[][their Table~1] {Duchesne2025}, G21: \citet[][radio component catalog]{Gordon2021}, G23: \citet{Gordon2023}, HW17: \citet{Hurley2017}, I17: \citet{Intema2017}, M03: \citet{Mauch2003}. For a larger compilation of major radio continuum surveys see \citet{Norris2017} and https://research.csiro.au/racs/home/survey/comparison/.}
\label{tab:surveys}
\end{table*}



\begin{acknowledgement}

We are grateful to Heinz Andernach for (a) his meticulous work on expanding the catalog of GRGs with large angular sizes to smaller angular sizes, presented in the Appendix, and (b) many fruitful discussions on GRG sizes, morphologies and environments. We also thank Ray Norris and Marcus Br\"uggen for comments on an earlier version of this paper. This scientific work uses data obtained from Inyarrimanha Ilgari Bundara / the Murchison Radio-astronomy Observatory. We acknowledge the Wajarri Yamaji People as the Traditional Owners and native title holders of the Observatory site. CSIRO’s ASKAP radio telescope is part of the Australia Telescope National Facility (\url{https://ror.org/05qajvd42}). Operation of ASKAP is funded by the Australian Government with support from the National Collaborative Research Infrastructure Strategy. ASKAP uses the resources of the Pawsey Supercomputing Research Centre. Establishment of ASKAP, Inyarrimanha Ilgari Bundara, the CSIRO Murchison Radio-astronomy Observatory and the Pawsey Supercomputing Research Centre are initiatives of the Australian Government, with support from the Government of Western Australia and the Science and Industry Endowment Fund. \\

The Legacy Surveys consist of three individual and complementary projects: the Dark Energy Camera Legacy Survey (DECaLS; Proposal ID \#2014B-0404; PIs: David Schlegel and Arjun Dey), the Beijing-Arizona Sky Survey (BASS; NOAO Prop. ID \#2015A-0801; PIs: Zhou Xu and Xiaohui Fan), and the Mayall z-band Legacy Survey (MzLS; Prop. ID \#2016A-0453; PI: Arjun Dey). DECaLS, BASS and MzLS together include data obtained, respectively, at the Blanco telescope, Cerro Tololo Inter-American Observatory, NSF’s NOIRLab; the Bok telescope, Steward Observatory, University of Arizona; and the Mayall telescope, Kitt Peak National Observatory, NOIRLab. The Legacy Surveys project is honoured to be permitted to conduct astronomical research on Iolkam Du’ag (Kitt Peak), a mountain with particular significance to the Tohono O’odham Nation. NOIRLab is operated by the Association of Universities for Research in Astronomy (AURA) under a cooperative agreement with the National Science Foundation. This project used data obtained with the Dark Energy Camera (DECam), which was constructed by the Dark Energy Survey (DES) collaboration. 

\end{acknowledgement}




\section*{Data availability} 

The ASKAP data used in this article are available through the CSIRO ASKAP Science Data Archive (CASDA\footnote{CASDA: \url{dap.csiro.au}}) under \url{https://doi.org/10.25919/5e5d13e6bda0c}. Additional data processing and analysis was conducted using the {\sc miriad} software\footnote{\url{https://www.atnf.csiro.au/computing/software/miriad/}} and the Karma visualisation\footnote{\url{https://www.atnf.csiro.au/computing/software/karma/}} packages. DES images were obtained through the Legacy Survey Viewer\footnote{\url{https://www.legacysurvey.org/viewer/}}. Combined ASKAP radio continuum images may be made available on reasonable request to the lead author after paper publication.


\printendnotes

\bibliography{pasa-grgs}

\clearpage

\appendix

\section*{Appendix A}

\begin{figure}
\centering
    \includegraphics[width=8cm]{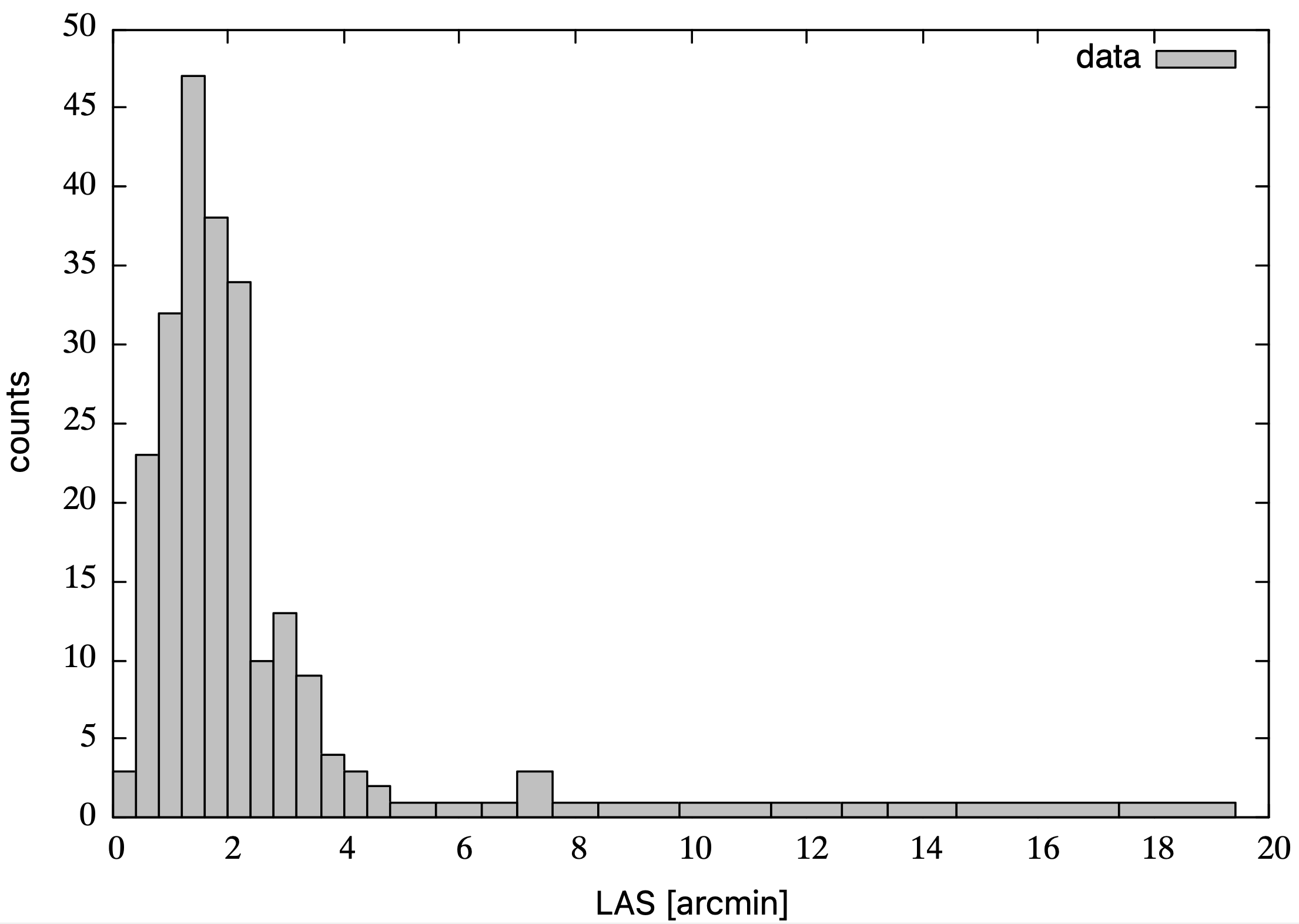} \\
    \includegraphics[width=8cm]{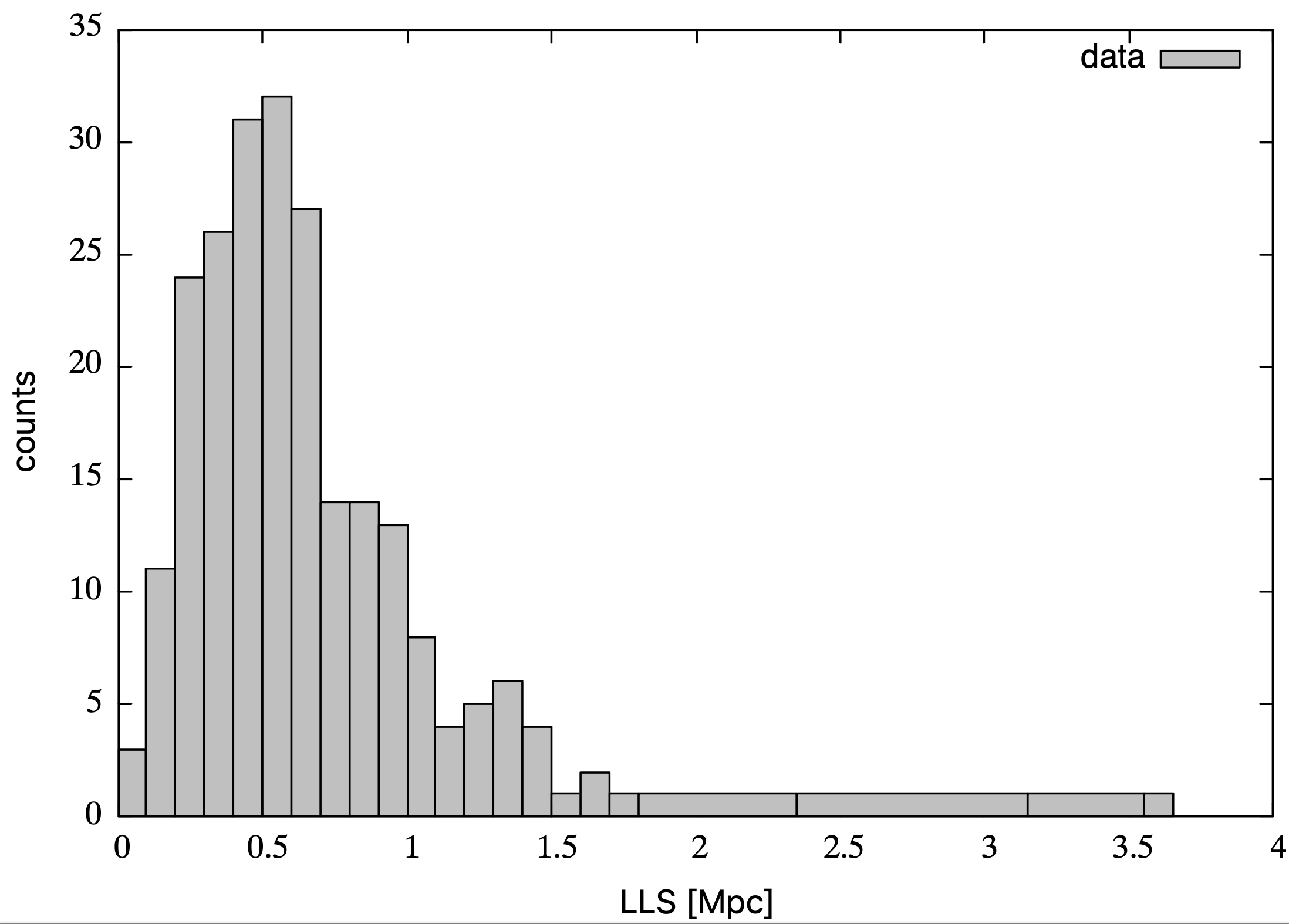}
\caption{Histograms of the LAS and LLS distributions of all radio galaxies catalogued in the ASKAP Sculptor field, as listed in Table~\ref{tab:appendix}.}
\label{fig:grg-las-lls}
\end{figure}

In Table~\ref{tab:appendix} we list the 232 radio galaxies catalogued in the $\sim$40~deg$^2$ ASKAP Sculptor field as well as two amorphous halos and ORC~J0102--2450. For LAS $\gtrsim$ 1.5\arcmin\ the sample is complete, but $\sim$100 smaller RGs are also included (see Fig.~\ref{fig:grg-las-lls}). For some of the latter, LAS was measured from the high-resolution VLASS 3~GHz images. Redshift estimates of the RG host galaxies were obtained from the literature as indicated. When available, we list the spectroscopic (s) redshift, otherwise the photometric (p) redshift from DES-DR9 \citep{Zhou2021} or the average photometric redshift from multiple references. In a few cases, we give our estimated (e) galaxy redshift. As new redshifts become available, the listed values may be superseded. At current count there are 35 radio galaxies $\ge$1~Mpc (incl. four candidates), 42 with sizes of 0.7 -- 1~Mpc (incl. two candidates), and 155 with sizes less than 0.7~Mpc (incl. two candidates). \\

\begin{table*}
\centering
\begin{tabular}{lclrlclcll}
\hline
ASKAP & LAS & $z$ & LLS  & Host Name & Type & & $z_{REF}$ & mag/filter & comments \\
Name & [arcmin] & & &  [Mpc] & & & \\
\hline
\hline
. \\
J0035--2557  & 1.39 & 0.81 & 0.63 & DES J003520.95--255707.7 & G & p & (1) & 22.29r & FR\,II \\
J0036--2607C & 6.05? & 0.05720 & 0.40 & 2MASX J00362661--2607564 & G & s & (2) & 16.02r & FR\,II remnant \\
J0036--2625  & 1.16 & 1.4 & 0.59 & DES J003608.26--262530.3 & QSOc & p & (3,4) & 18.88r & FR\,II, complex, asym. \\
J0037--2415  & 1.57 & 0.63 & 0.64 & DES J003728.86--241539.1 & G & p & (1) & 21.51r & FR\,II, DDRG \\
J0037--2707  & 2.27 & 0.98 & 1.09 & DES J003714.11--270726.9 & G & p & (1) & 22.35r & FR\,II \\
{\bf J0037--2752}  & {\bf 7.5} & 0.2389 & {\bf 1.70} & WISEA J003716.97--275235.3 & G & s & (2) & 17.77rK & {\bf FR\,II} \\
J0038--2336  &  1.82  & 0.66 & 0.76 &  DES J003854.22--233604.1 & G & p & (1,17) & 21.69r & FR\,I, WAT? \\
J0038--2342  & 1.4 & 0.21 & 0.29 & SDSS J003858.10--234207.1 & G & p & (1,8,11,17) & 17.82r & FR\,II remnant, WAT?  \\
J0038--2414 & $>$1.5 & 0.325 & $>$0.42 & DES J003837.46--241414.6 & G & p & (1,8) & 18.37r & FR\,I \\
J0038--2459 & 1.58 & 0.498 & 0.58 & DES J003814.72--245902.0 & QSOc & s & (7) & 19.05r & FR\,I/II \\ \\

J0038--2504 & 2.0 & 0.06292 & 0.15 & WISEA J003800.32--250456.6 & G & s & (2) & 14.85r & FR\,I/II (in A2800) \\
J0038--2508  & 1.1 & 0.0730 & 0.09 & 2MASX J00385303--2508543 & G & s & (7) & 14.96r & FR\,I, twin jet? \\
J0038--2626 & 1.08 & 1.2? & 0.54 & DES J003820.76--262639.9 & G & e & --- & 23.16r & FR\,II, no core \\
J0039--2425 & 0.95 & 1.01 & 0.97 & DES J003906.57--242517.6 & G & p & (1,9) & 21.34r & FR\,II, bent \\
{\bf J0039--2541} & 15.5 & 0.07297 & 1.29 & WISEA J003930.86--254147.8 & G & s & (2) & 14.88r & {\bf FR\,I} \\
J0039--2612 & 1.35 & 0.11094 & 0.16 & 2MASX J00395528--2612433 & G & s & (7) & 15.45r & FR\,I, NAT (in A88) \\
J0039--2752  & 0.8 & 0.8339 & 0.37 & DES J003929.62--275235.5 & QSO & s & (5) & 19.28r & FR\,II, asym. \\
J0039--2800  & 0.7 & 0.846 & 0.32 & DES J003911.11--280056.0 & G & p & (1) & 21.79r & DDRG, FR\,II \\
J0040--2524 & 0.75 & 1.0? & 0.36 & DES J004052.86--252446.2 & QSOc & e & -- & 22.66r & FR\,II \\ 
J0040--2526 & 1.73 & 0.880 & 0.86 & WISEA J004022.45--252654.1 & G & p & (1) & 23.23r & FR\,II \\ \\

J0040--2530 & $>$3.05 & 0.06340 &  $>$0.22 & WISEA J004049.34--253023.9 & G & s & (12) & 15.61r & FR\,I (in A2800) \\
J0040--2623  &  2.07  & 0.11243 & 0.25 & DES J004039.62--262342.8 & G & s & (2) & 17.13r & FR\,II remnant \\
J0040--2743  & 1.0 & 0.78 & 0.45 & DES J004019.33--274328.5 & G & p & (8) & 21.57r & FR\,II relic \\
J0041--2530 & 0.45 & 0.63 & 0.18 & WISEA J004104.87--253013.8 & G & p & (1,8) & 20.14r & FR\,I, bipolar jet \\
{\bf J0041--2655} & {\bf 7.3} & 0.232 & {\bf 1.62} & WISEA J004119.25--265548.3 & G & p & (1) & 18.17r & {\bf FR\,II, relic lobes} \\
J0041--2656 & 0.8 & 0.67 & 0.34 & DES J004117.66--265611.9 & G & p & (1) & 21.08r & FR\,I/II remnant \\
J0041--2657 & 0.57 & 1.0 & 0.27 & DES J004102.87--265729.5 & G & p & (1) & 23.97r & FR\,II bent, no core \\
J0041--2812 & 2.22 & 0.225 & 0.48 & DES J004122.69--281229.7 & G & p & (1,8) & 17.77r & FR\,II \\
J0042--2143 & 1.3 & 0.19 & 0.25 & DES J004202.82--214342.3 & G & p & (1) & 17.27r & FR\,I/II \\
J0042--2242  & 1.9 & 0.45 & 0.66 & DES J004204.60-224222.3 & G & p & (1) & 19.13r & FR\,II, WAT? \\ \\

J0042--2243  & 4.0 & 0.42? & 1.38 & DES J004222.68--224304.2 ? & G & p & (1,11,17) & 21.31r' & FR\,II \\
J0042--2332  & 1.7 & 0.704 & 0.73 & DES J004239.67--233224.3 & G & p & (9) & 20.42r & FR\,II \\
J0042--2409C & 2.36 & 0.51 & 0.87 & DES J004225.44--240940.1 & G & p & (1) & 20.37r & FR\,II relic? \\
J0042--2650  & 2.03 & 0.225 & 0.44 & DES J004224.43--265041.4 & G & p & (1,8) & 17.81r  & FR\,I, WAT? \\
J0042--2729 & 0.6 & 0.492 & 0.22 & PSO J004258.321--272917.33 & G & p & (8) & 20.00rK & FR\,I/II relic \\
J0043--2236 & 1.01 & 1.03 & 0.48 & DES J004340.61-223607.3 & G? & p & (1) & 23.57r & FR\,II \\
J0043--2258  & 1.9 & 0.41 & 0.62 & DES J004300.51--225825.9 & G & p & (1,8,11,17) & 19.49r & FR\,I/II remnant \\
J0043--2402 & 0.76 & 1.25 & 0.38 & DES J004316.73--240255.9 & G? & p & (1) & 24.33r & FR\,II \\
J0043--2511  & 2.0 & 0.345 & 0.59 & DES J004304.31--251140.9 & G & p & (1,8) & 18.55r & precess., FR\,II relic \\
J0043--2558 & 0.97 & 0.63 & 0.40 & DESI J10.9587--25.9833 & G? & p & (1) & 21.51r & FR\,II, no core \\
\hline
\end{tabular}
\caption{Properties of radio galaxies in the $\sim$40~deg$^2$ ASKAP Sculptor field. Bold source names denote the GRGs already presented in Table~1; the "C" in Col.~(1) stands for candidate. ---  References: (1) \citet[DES-DR9][]{Zhou2021}, (2) \citet{Colless2001}, 
(3) \citet{Katgert1998},
(4) \citet{Flesch2015}, (5) \citet{Croom2004}, (6) \citet{Owen1995}, (7) \citet{Jones2009}, (8) \citet[][p]{Bilicki2016}, (9) \citet{Zou2019}, (10) \citet{Pocock1984}, (11) \citet{Brescia2014}, (12) \citet{Way2005}, (13) \citet{Vettolani1989}, (14) \citet{Kirshner1983}, (15) \citet{Baker1995}, (16) \citet{Krogager2018}, (17) \citet[][SDSS-DR16]{Ahumada2020}, (18) \citet{Brown2001}, (19) \citet{Collins1995} --- Redshift coding: spectroscopic (s), photometric (p), our own photometric estimate (e). Magnitude coding letters indicate the band.}
\label{tab:appendix}
\end{table*}

\begin{table*}
\centering
\begin{tabular}{lclrlclcll}
\hline
ASKAP & LAS & $z$ & LLS  & Host Name & Type & & $z_{REF}$ & mag/filter & comments \\
Name & [arcmin] & & [Mpc] & & & \\
\hline
... continued \\

J0043--2626  & 1.52$^1$ & 0.74 & 0.67 & DES J004348.33--262626.8 & G & p & (1) & 21.91r  & FR\,II \\
J0044--2133 & 0.55 & 0.38 & 0.17 & PSO J004455.753--213340.90 & G & p & (8) & 21.04r' & FR\,I/II \\
J0044--2212 & 0.9 & 0.47 & 0.32 & DES J004411.01--221227.2 & QSOc & p & (1) & 23.15r & FR\,II, complex \\
J0044--2314 & 2.0 & 0.15 & 0.31 & DES J004426.05--231405.1 & G & p & (1) & 17.68r & FR\,I \\
{\bf J0044--2317C} & {\bf 6.0?} & 0.362 & {\bf 1.82} & WISEA J004426.72--231745.8 & G & p & (1) & 18.78r & {\bf HyMoRS} \\
J0044--2452  & 2.6 & 0.245 & 0.68 &  DES J004435.81--245202.9 & G & p & (1) & 17.94r  & FR\,II relic \\
J0044--2506 & 0.45 & 1.008 & 0.22 & DES J004442.98--250606.4 & G & p & (9) & 22.67r & FR\,II \\
J0044--2602 & 1.13 & 0.5 & 0.41 & PSO J004404.648--260231.65 & G & p & (8) & 19.72r & FR\,II relic, no core \\
J0044--2710  &  1.55 & 0.85  & 0.71 & DES J004426.68--271032.5 & G & p & (1) & 21.83r & FR\,II \\
J0045--2204  &  1.8  & 1.04  & 0.87 & DES J004538.60--220439.1 & G & p & (1) & 21.97r & FR\,II, complex \\ \\

J0045--2239 & 1.15 & 0.215 & 0.24 & SDSS J004523.93--223914.3 & G & p & (1,8,11,17) & 17.90r & FR\,II relic, WAT? \\
J0045--2411 & 1.02 & 0.79 & 0.46 & DES J004503.06--241141.1 & G & p & (1) & 21.26r & FR\,II relic \\
J0045--2434 & 1.27 & 0.807 & 0.57 & DES J004502.95--243423.9 & G & s & (10) & 21.2 & FR\,II \\
J0045--2501 & 3.5 & 0.11035 & 0.42 & WISEA J004506.98--250147.0 & G & s & (2) & 16.07r & spiral DRAGN \\
J0045--2753  &  1.88  & 0.62 & 0.77 &  DES J004518.34--275358.3 & G & p & (1) & 20.73r  & FR\,II relic \\
J0046--2219  & 1.4 & 0.73 & 0.61 & DES J004634.44--221933.0 & G & p & (1) & 20.65r & FR\,II remnant \\
J0046--2243  & 1.62 & 0.42 & 0.54 & SDSS J004651.05--224303.0 & G & p & (1,8,11,17) & 19.73r' & FR\,II, no core \\
J0046--2324  & $>$2.0 & 0.23 & $>$0.44 & DES J004643.28--232416.6 & G & p & (1,8) & 17.55r & FR\,I/II  \\
J0046--2554  &  2.86  & 0.93 & 1.35 &  DES J004600.79--255400.2 & G & p & (1) & 22.72r & FR\,II, DDRG \\
J0046--2703  & 0.9 & 1.02 & 0.43 & DES J004613.54--270336.0 & G & p & (1) & 23.79r & FR\,II \\ \\

J0046--2733  & 1.47 & 0.35 & 0.44 & DES J004658.54--273302.8 & G & p & (1) & 18.05r & FR\,I/II \\
J0046--2755  & 1.34 & 0.57 & 0.52 & DES J004615.34--275501.5 & G & p & (1,8) & 19.20r & FR\,I/II \\
{\bf J0047--2419}  & {\bf  5.0} & 0.27 & {\bf 1.24} & WISEA J004709.94--241939.6 & G/QSO? & p & (1) & 17.47r & {\bf FR\,II relic} \\ 
J0047--2453 & 1.3 & 0.42 & 0.43 & DES J004743.84--245345.2 & G & p & (1,8) & 19.95r & FR\,II \\
J0047--2657 & 1.39 & 0.93 & 0.66 & DES J004757.58--265756.0 & G & p & (1) & 22.76r & FR\,II, WAT? \\
J0047--2725  & 2.8? & 1.057 & 1.36 &  DES J004704.71--272545.0 & QSO & s & (5) & 20.70r & FR\,II relic \\
J0047--2731 & 0.62 & 0.545 & 0.24 & DES J004722.94--273110.2 & G & p & (1,18) & 19.29r & FR\,II, WAT? \\
J0048--2152 & $>$4.34 & 0.22 & 0.92 & DES J004858.74--215208.9 & G & p & (1,8,9) & 18.97r & FR\,II relic, WAT\\
J0048--2527 & 1.6 & 0.212 & 0.33 & DES J004801.33--252738.1 & G & p & (1) & 17.45r & FR\,I/II, WAT? \\
J0048--2729 & 1.25 & 0.35 & 0.37 & DES J004836.13--272942.1 & G & p & (1) &  18.79r & FR\,II \\ \\

J0048--2804 & 2.01 & 0.64 & 0.83 &  DES J004822.93--280419.6 & G & p & (1) & 21.61r & FR\,II, complex \\
{\bf J0049--2137} & {\bf 7.23} & 0.23 & {\bf 1.59} & DES J004941.58--213722.1 & G & p & (1) & 17.91r & {\bf FR\,II remnant} \\
J0049--2203 & 3.3 & 0.19591 & 0.64 & 2MASX J00490997--02203567 ? & G & s & (7) & 16.52Rc & FR\,II \\
J0049--2314  & 2.02 & 0.79 & 0.91 & DES J004950.56--231413.3 & G & p & (1) & 20.96r & FR\,I \\
J0049--2347 & 2.9 & 0.12 & 0.38 & 2MASX J00492151--2347017 & G & p & (1) & 16.74r & FR\,I, WAT? \\
J0049--2417  & 1.14 & 1.15 & 0.56 & DES J004941.54--241732.9 & ? & p & (1) & 23.96r & FR\,II \\
J0049--2555 & 1.04 & 1.06 & 0.51 & DES J004901.91--255517.1 & G & p & (1) & 22.93r & FR\,II \\
J0049--2556 & 1.83 & 0.77 & 0.81 & DES J004956.44--255609.5 & G & p & (1) & 21.41r & FR\,II \\
{\bf J0050--2135} & {\bf 18.0} & 0.0576 & {\bf 1.20} & WISEA J005046.49--213513.6 & G & s & (7) & 14.11r & {\bf FR\,I, restarting} \\
J0050--2143  & $>$2.0? & 0.09520 & $>$0.21 & 2MASX J00500665--2143241 & G & s & (7) & 15.88r & one-sided \\
\hline 
\end{tabular}
\end{table*}

\begin{table*}
\centering
\begin{tabular}{lclrlclcll}
\hline
ASKAP & LAS & $z$ & LLS  & Host Name & Type & & $z_{REF}$ & mag/filter & comments \\
Name & [arcmin] & & [Mpc] & & & \\
\hline
... continued \\

J0050--2152  & 2.22 & 0.06251 & 0.16 & 2MASX J00505686--2152167 & G & s & (14) & 16.07r & FR\,I/II, HyMoRS?\\ 
J0050--2206  & 0.75 & 0.662 & 0.31 & DES J005048.25--220653.0 & G & p & (9) & 21.60r & FR\,II, no core \\
J0050--2211  & 1.87 & 0.42 & 0.62 & DES J005052.39--221131.3 & G & p & (1) & 18.18r & FR\,II relic \\
{\bf J0050--2325} & {\bf 13.5} & 0.1114 & {\bf 1.60} & WISEA J005049.89--232511.1 & G & s & (7) & 15.70r & {\bf FR\,II, WAT} \\ 
J0050--2346 & 2.46 & 1.07 & 1.20 & DESI J012.6117--23.7719 & G & p & (9) & 23.76r & FR\,II \\ 
J0050--2353  & 1.47 & 0.64 & 0.61 & DES J005014.18--235337.3 & G & p & (1) & 21.03r & FR\,II \\
J0050--2408  & 1.95 & 0.96 & 0.93 & DES J005015.26--240832.0 & G & p & (1) & 23.01r & FR\,II \\
J0050--2409  & 1.84 & 0.52 & 0.69 & DES J005004.72--240943.9 & G & p & (1) & 20.03r & FR\,II, twin jet \\
J0050--2446 & 1.63 & 0.97 & 0.78 & DES J005004.96--244600.5 & G & p & (1) & 23.55r & FR\,II \\
J0050--2454  & 1.36 & 0.83 & 0.62 & DES J005015.17--245438.6 & G & p & (1) & 21.52r & FR\,II, WAT? \\ \\

J0050--2541  & 1.42 & 0.77736 & 0.63 & DES J005040.94--254122.9 & QSO & s & (7)  & 18.72r & FR\,II, core-dominated \\
J0051--2104 & 1.97 & 0.85 & 0.91 & DES J005153.66--210420.5 & G & p & (1) & 23.31r & FR\,II remnant \\
J0051--2256  & 0.6 & 0.740 & 0.26 & DES J005142.11-225638.0 & G? & p & (9) & 22.65r & FR\,II remnant \\
J0051--2420 & 1.97 & 1.5? & 1.00 & DESI J012.8370--24.3465 & QSOc & e  & (1) & $>$24r & FR\,II \\
J0051--2434 & 1.62 & 1.07 & 0.79 & DES J005109.22--243449.1 & G & p & (1) & 24.67r & FR\,II \\
J0051--2459 & 1.24 & 0.93 & 0.58 & DES J005112.88--245906.0 & G & p & (1) & 21.71r & FR\,II \\
J0051--2531  & 1.63 & 0.61 & 0.66 & DES J005111.27--253125.2 & G & p & (1) & 21.42r & FR\,II \\
J0051--2626  & 0.83 & 0.84 & 0.37 & DES J005105.23--262639.1 & G & p & (1) & 22.62r & FR\,II \\
J0051--2720  & 0.86 & 0.43 & 0.29 & DES J005116.35--272046.0 & G & p & (1,8,18) & 19.00r & FR\,II \\
J0052--2425  & $>$2.9 & 0.19 & $>$0.55 & DES J005238.97--242500.9 & G & p & (1,8) & 17.49r & FR\,I, WAT \\ \\

J0052--2532  & 2.36 & 0.74 & 1.03 & DES J005254.18--253256.0 & G & p & (1) & 21.85r & FR\,II relic \\
J0052--2653 & 2.93 & 0.68 & 1.24 & DES J005246.78--265308.6 & G & p & (1) & 20.85r & FR\,I, WAT? \\
J0052--2727  & 1.49 & 0.80 & 0.67 & DES J005237.56--272719.7 & G & p & (1) & 21.93r & FR\,I/II \\
J0053--2144 & 3.1 & 0.0593 & 0.21 & 2MASX J00532556--2144117 & G & s & (6) & 15.29rK & FR\,I/II (in A114) \\
J0053--2408 & 1.3 & 0.43 & 0.44 & DES J005339.35--240856.5 & G & p & (1,8) & 18.37r & FR\,II \\
J0053--2531  & 1.75 & 0.53 & 0.66 & DES J005358.09--253147.1 & G & p & (1) & 20.00r & FR\,I, WAT \\
J0053--2649  & 1.28 & 1.05 & 0.62 & DES J005342.19--264907.6 & G & p & (1) & 23.64r & FR\,II \\
J0053--2816C & 3.21 & 0.55? & 1.15 & DES J005304.85--281613.6 ? & G & p & (1,18) & 21.76r & FR\,II, uncertain \\
J0054--2124  & 1.82 & 0.48 & 0.65 & DES J005434.46--212452.4 & G & p & (1) & 19.47r & FR\,II \\
J0054--2322  & 1.67 & 0.22 & 0.36 & 2MASX J00542023--2322245 & G & p & (1) & 17.21r & FR\,II, BCG \\ \\

J0054--2519  & 2.37 & 0.51 & 0.88 & DES J005405.26--251924.6 & G & p & (1) & 20.48r & FR\,II \\
J0054--2657  & 1.36 & 0.50? & 0.50 & DES J005400.91--265727.6 & G & p & (1,8) & 19.51r & FR\,I/II relic, bent \\
J0054--2733 & 1.58 & 1.27 & 0.79 & DES J005429.64--273333.2 & QSOc & p & (1) & 23.84r & FR\,II \\
J0055--2217 & 2.7 & 0.23 & 0.59 & DES J005559.85--221710.4 & G & p & (1) & 17.13r & FR\,II remnant, core-dominated \\
{\bf J0055--2231} & {\bf 9.2} & 0.11437 & {\bf 1.14} & WISEA J005548.98--223116.9 & G & s & (7) & 16.10rK & {\bf FR\,I} \\
J0055--2306  & $>$2.06 & 0.62 & $>$0.84 & DES J005547.33--230647.0 & G & p & (1) & 20.63r & FR\,II, asym. \\
J0055--2308  & 1.08 & 1.25 & 0.54 & DES J005552.69--230844.7 & G/QSOc & p & (3,4) & 18.58r & FR\,II \\
J0055--2318  & 1.44? & 0.24 & 0.33 & DES J005506.61--231804.7 & G & p & (1,8) & 18.53r & FR\,I, WAT, asym. \\
J0055--2409 & 1.92 & 0.98 & 0.92 & DES J005556.90--240913.4 & G & p & (1) & 22.71r & FR\,I \\
J0055--2544  & 1.22 & 0.435 & 0.41 & DES J005528.83--254441.6 & G & p & (1,8) & 19.78r & FR\,II relic \\
\hline
\end{tabular}
\end{table*}

\begin{table*}
\centering
\begin{tabular}{lclrlclcll}
\hline
ASKAP & LAS & $z$ & LLS  & Host Name & Type & & $z_{REF}$ & mag/filter & comments \\
Name & [arcmin] & & [Mpc] & & & \\
\hline
... continued \\

J0055--2621 & 3.6 & 0.11585 & 0.45 & WISEA J005550.06--262155.9 & G & s & (19) & 15.67r & HT (in Abell~118) \\
J0055--2647  & 1.48 & 0.836 & 0.68 & DES J005517.17--264756.1 & G & p & (9) & 22.08r & FR\,II, no core \\
J0055--2647 & 1.48 & 0.836 & 0.68 & DES J005517.17--264756.1 & G & p & (9) & 22.08r & FR\,II, no core \\
J0055--2746  & 2.22 & ? & & VLASS QL J005556.57--274623.6 & ? & & & & FR\,II \\
J0055--2752  & $>$0.49 & 0.61 & 0.20 &  DES J005513.28--275207.6 & G & p & (1) & 20.42r & FR\,II \\
J0056--2110 & 2.46 & 0.235 & 0.55 & 2MASX J00562465--2110121 & G & p & (1,8) & 17.40r & FR\,I/II, BCG, WAT \\
J0056--2237  & 1.43 & 0.224 & 0.31 & DES J005639.58--223747.9 & G & p & (1) & 17.65r & FR\,II relic \\
J0056--2306  & 3.32 & 0.42 & 1.10 & DES J005635.64--230616.0 & G & p & (1) & 20.19r & FR\,II \\
J0056--2359  & 1.3 & 0.205 & 0.26 & DES J005650.93--235900.2 & G & p & (1,8) & 18.16r & FR\,I/II, WAT? \\
J0056--2418  & 1.4 & 0.88 & 0.65 & DES J005644.22--241853.2 & G & p & (1) & 22.83r & FR\,II remnant \\ \\

J0056--2625a & 2.07 & 0.23 & 0.46 & DES J005603.45--262502.7  & G & p & (1) & 17.83r & FR\,I/II \\
J0056--2625b  & 0.51 & 0.50 & 0.19 & DES J005608.16--262505.1 & G & p & (1,8) & 20.27r & FR\,II, no core \\
J0056--2711 & 2.86 & 0.50 & 1.05 & DES J005643.54--271100.2 & G & p & (1) & 19.71r & FR\,I/II, HyMoRS, WAT asym.? \\
J0057--2311  & 0.2 & 0.96 & 0.01 & DES J005718.70--231125.0 & G & p & (1) & 22.67r & FR\,II relic \\
J0057--2311C & 4.11 & $>$0.47 & $>$1.45 & DES J005725.22--231137.8 ? & G & p & (1,8) & 23.01r & FR\,I, DDRG? \\
{\bf J0057--2428}  & {\bf 12.1} & 0.25 & {\bf 2.84} & DES J005736.29--242814.8 & G & p & (1,8) & 18.90r & {\bf FR\,II}
\\
J0057--2506a & 0.77 & 0.60 & 0.31 & DES J005741.78--250650.2 & G & p & (1) & 20.23r & FR\,II \\
J0057--2506b & 2.91 & 0.20100 & 0.58 & 2MASX J00575181--2506128 & G & s & (2) & 17.43r & FR\,I, WAT, twinjet \\
J0057--2527 & 2.3 & 0.31 & 0.63 & DES J005726.91--252742.1 & G & p & (1) & 19.20r & FR\,I, WAT \\
J0057--2533  & 1.02 & 0.54 & 0.39 & DES J005742.75--253315.3 & G & p & (1,8) & 19.99r & FR\,II \\ \\

J0057--2542 & 1.90 & 0.90 & 0.89 & DES J005715.31--254211.1 & G & p & (1) & 22.36r & FR\,II \\
J0057--2616 & 2.6 & 0.11281 & 0.31 & WISEA J005722.95--261651.6 & G & s & (7) & ? & A122 (BCG lobes?) \\
J0057--2643 & 2.3 & 0.37 & 0.71 & DES J005745.57--264320.8 & G & p & (1,8) & 19.33r & FR\,I, WAT \\
J0058--2005  & 1.6 & 0.9 & 0.75 & DES J005800.99--200558.2 ? & G & p & (1) & 23.13r & FR\,II \\
J0058--2013  & 1.26 & 0.52 & 0.47 & DES J005828.79--201316.2 & G & p & (1) & 20.29r & FR\,II relic \\
J0058--2125  & 0.58 & 0.5157 & 0.22 & PSO J005838.528--212508.77 & QSO & s & (7) & 18.53rK & FR\,II \\
J0058--2310 & 3.1 & 0.37 & 0.95 & DES J005857.47--231033.8 & G & p & (1,8) & 19.46r & FR\,II relic \\
J0058--2324 & 4.63 & 0.195 & 0.90 & 2MASX J00583393--2324166 & G & p & (1,8,11) & 16.30r & FR\,I, DDRG, WAT?, core-dom. \\
J0058--2401 & 1.75 & 0.35 & 0.52 & DES J005827.81--240104.6 & G & p & (9) & 18.95r & FR\,II remnant \\
J0058--2518  & 0.85 & 0.325 & 0.24 & DES J005820.96--251826.3 & G & p & (1) & 18.14r & FR\,I/II ?, WAT? \\ \\

J0058--2529  & 1.77 & 0.59 & 0.70 & DES J005819.39--252940.5 & G & p & (1) & 20.47r & FR\,II, complex \\
{\bf J0058--2625}  & {\bf 12.0} & 0.11341 & {\bf 1.48} & 2MASX J00583576--2625214 & G & s & (2) & 16.38r & {\bf FR\,I/II}, WAT? twin jet \\
J0059--2105 & 2.57 & 1.16 & 1.27 & DES J005947.01--210510.1 ? & G & p & (1) & 24.08r & FR\,II, no core\\
J0059--2139  & 0.55 & 0.932 & 0.26 & DES J005918.55--213946.8 & G /QSO? & p & (9) & 21.34r & complex \\
J0059--2149  & 0.4 & 0.553 & 0.15 & DES J005916.01--214902.0 & G/QSO? & p & (9) & 22.66r & FR\,II \\
J0059--2205C & 2.82 & 0.79 & 1.27 & DES J005911.87--220511.9 & G & p & (1) & 20.88r & FR\,II, asym. ? \\
J0059--2240 & 2.4? & 0.483 & 0.86 & DES J005952.23--224016.4 & G & p & (1,8) & 19.35r & FR\,I \\
J0059--2246  & 1.66 & 0.29 & 0.43 & DES J005932.19--224604.7 & G & p & (8,9) & 17.66r & FR\,II remnant \\
{\bf J0059--2352C} & {\bf 7.96} & 0.735 & {\bf 3.49} & WISEA J005954.72--235254.7 & G & p & (1) & 22.44r & {\bf FR\,II} \\
J0059--2434 & 1.21 & 0.92 & 0.57 & DES J005953.53--243452.6 & G & p & (1) & 22.52r & FR\,II relic \\
\hline
\end{tabular}
\end{table*}

\begin{table*}
\centering
\begin{tabular}{lclrlclcll}
\hline
ASKAP & LAS & $z$ & LLS  & Host Name & Type & & $z_{REF}$ & mag/filter & comments \\
Name & [arcmin] & & [Mpc] & & & \\
\hline
... continued \\

J0059--2456 & 2.18 & 0.215  & 0.46 & DES J005953.78--245625.6 & G & p & (1) & 17.77r & FR\,I/II, WAT? \\
J0059--2540 & 3.09 & 0.34 & 0.90 & DES J005941.51--254003.0 & G & p & (1,8) & 18.08r & FR\,I/II precess. \\
J0059--2646 & 3.9 & 0.345 & 1.14 & DES J005948.78--264639.2 & G & p & (1,8) & 18.53r & FR\,I/II, twin jet \\ 
J0059--2813  & 1.70 & 0.22 & 0.36 & DES J005926.97--281328.5 & G & p & (1) & 17.96r & FR\,II \\
J0100--2012  & 2.4  & 0.46  & 0.84 & DES J010057.11--201222.4 & G & p & (1,8) & 19.96r & FR\,II relic \\
{\bf J0100--2125} & {\bf 6.34} & 0.193 & {\bf 1.21} & 2MASX J01003900--2125340 & G & p & (1) & 16.89r & {\bf FR\,I} \\
J0100--2137  & 1.37 & 0.297 & 0.36 & 2MASX J01000664--2137044 & G & p & (8) & 18.26rK & FR\,II remnant, bent \\
J0100--2239  & 1.07 & 0.706 & 0.46 & PSO J010042.098--223959.97 & G/QSO? & s & (15) & 29.39rK & FR\,II plume \\
J0100--2446 & 2.11 & 0.85 & 0.97 & DES J010033.36--244654.9 ? & G & p & (1) & 23.05r & FR\,II \\
J0100--2455  & 1.33 & 0.37 & 0.41 & DES J010028.30--245514.8 & G & p & (1,8) & 19.20r & FR\,I/II, precess. \\ \\

J0100--2511  & 1.22 & 0.24 & 0.25 & DES J010037.02--251114.4 & G & p & (1) & 17.44r & FR\,II \\
J0100--2600 & 2.4 & 0.315 & 0.66 & DES J010045.68--260054.1 & G & p & (1,8) & 18.15r & WAT, FR\,I \\
J0100--2701a & 1.04 & 0.97 & 0.50 & DES J010035.64--270151.3 & G/QSO? & p & (1) & 21.94r & FR\,II \\
J0100--2701b & 1.65? & 1.5? & 0.84 & CWISE J010058.44--270156.2 & QSOc & e & -- & 16.52W2 & FR\,II, asym. \\
J0100--2814  & 1.12 & 1.17 & 0.56 & DES J010046.03--281425.8 & G & p & (1) & 25.05r & FR\,II \\
J0101--2044  & 2.37 & 0.295 & 0.63 & DES J010126.78--204443.4 & G & p & (1,8) & 18.53r & amorphous halo, end-on? \\
J0101--2600  & 1.48 & 0.62 & 0.60 & DES J010111.13--260005.4 & G & p & (1) & 20.27r & FR\,II relic \\
J0101--2604  & 2.15 & 1.00  & 1.03 & DES J010108.98--260400.3 & G & p & (1) & 23.46r & FR\,II \\
J0101--2752  & 0.92 & 0.385 & 0.29 & DES J010143.41--275208.1 & G & p & (1) & 19.44r & FR\,II relic \\
J0102--2018C & 2.55 & 0.335 & 0.73 & DES J010230.83--201855.1 & G & p & (1,8) & 20.00r & FR\,II \\ \\

J0102--2154a & $>$0.1 & 0.293 & 0.03 & PSO J010245.191--215415.2 & GPair & s & (6) & 17.89rK & FR\,II ? \\
{\bf J0102--2154b}  & 6.38 & 0.293 & 1.68 & 2MASX J01024529--2154137 & GPair? & s & (6) & 18.57rP &  {\bf FR\,I/II relic, precess.} \\
J0102--2248  & 0.56 & 0.3? & 0.15 & PSO J010257.3960--224812.153 & QSOc & e & -- & 18.81rK & FR\,II remnant \\
J0102--2413  & 0.6 & 0.487 & 0.22 & DES J010209.39--241340.0 & G & p & (1) & 18.83r & FR\,II relic \\
J0102--2450  & 2.33 & 0.27 & 0.58 & DES J010224.33--245039.5 & G & p & (9) & 18.97r & ORC J0102--2450 \\
J0102--2451  & 0.58 & 0.31 & 0.16 & DES J010213.03--245151.1 & G & p & (8,9) & 19.38r & FR\,I, HT \\
J0102--2523  & 3.53 & 0.285 & 0.91 & DES J010201.41--252325.3 & G & p & (1,8) & 19.50r & FR\,II relic \\
J0102--2554  & 1.34 & 1.5? & 0.68 & WISEA J010208.31--255422.6 & G & e & -- & 16.27W2 & FR\,II \\
J0102--2623  & 1.61 & 1.13  & 0.79 & DES J010233.73--262306.6 & G & p & (1) & 22.66r & FR\,II \\
J0102--2640  & 1.83 & 0.86 & 0.84 & DES J010218.47--264043.9 & G & p & (1) & 23.66r & FR\,II \\ \\

J0102--2755  & 1.41 & 0.50 & 0.52 & DES J010247.94--275521.0 & G & p & (1) & 20.32r & FR\,II relic \\
J0103--2100  & 3.55 & 0.363 & 1.08 & DES J010304.74--210026.4 & G & p & (1,8) & 19.12r & FR\,II relic \\
J0103--2133  & 1.38 & 0.22 & 0.29 & 2MASX J01032091--2133180 & G & p & (1) & 17.46r & FR\,II relic \\
J0103--2200  & 2.27 & 0.27 & 0.56 & 2MASX J01033676--2200060  &  G & p & (1,8,11) & 17.48r & FR\,II remnant \\
J0103--2236  & 1.92 & 0.225 & 0.42 & 2MASX J01033583--2236060 & G & p & (1,8) & 17.61r & FR\,II \\
J0103--2330  & 2.57 & 0.16 & 0.43 & 2MASX J01032624--2330120 & G & p & (1,8,11) & 16.69r & FR\,I/II, HyMoRS? \\
J0103--2439C & 1.47 & 1.5? & 0.75 & WISEA J010332.22--243926.5 & ? & e & -- & 16.18W2 & FR\,II, no core \\
J0103--2514  & 3.81 & 1.07 & 1.85 & DES J010327.91--251446.9 & G & p & (1) & 24.15r & FR\,II \\
J0104--2034  & 2.11 & 0.52 & 0.79 & DES J010430.98--203432.2 & G & p & (1) & 20.62r & FR\,II \\
J0104--2156  & 0.42 & 0.870 & 0.19 & DES J010447.18--215625.5 & G & p & (9) & 21.96r & FR\,II, no core \\ 
\hline
\end{tabular}
\end{table*}

\begin{table*}
\centering
\begin{tabular}{lclrlclcll}
\hline
ASKAP & LAS & $z$ & LLS  & Host Name & Type & \multicolumn{2}{c}{$z_{REF}$} & mag/filter & comments \\
Name & [arcmin] & & [Mpc] & & & \\
\hline
... continued \\

J0104--2207  & 0.9? & 0.39 & 0.29 & PSO J010451.704-220706.50 & G & p & (8) & 18.89rK & complex, WAT? \\
J0104--2244  & 1.36 & 0.56 & 0.53 & DES J010433.51--224422.6 & G & p & (1) & 19.73r & FR\,II \\
J0104--2517C & 1.65 & 1.5? & 0.84 & WISEA J010457.71--251703.9 & ? & e & -- & 17.16W2 & FR\,II ?\\
J0105--2146  & 0.98 & 1.7? & 0.50 & PSO J010545.634--214657.60 & QSOc & s & (7) & 18.30rK & FR\,II, asym. \\
J0105--2344  & 2.78 & 0.41 & 0.91 & DES J010535.39--234440.6 & G & p & (1,8) & 19.50r & FR\,II remnant \\
J0105--2347  & 2.15 & 0.236 & 0.48 & DES J010520.68--234721.9 & G & p & (1) & 17.32r & FR\,I/II, WAT? \\
J0105--2419  & 1.25 & 0.265 & 0.31 & DES J010519.56--241956.3 & G & p & (1,8) & 17.84r & one-sided WAT? \\
J0105--2543  & 0.55 & 0.90 & 0.26 & DES J010537.56--254342.9 & G & p & (1) & 22.60r & FR\,I relic ? \\
J0105--2547  & 0.19 & 0.33 & 0.05 & DES J010545.20--254751.6 & G & p & (1,8,9) & 17.58r & FR\,II \\
J0105--2603  & 1.2 & 1.0 & 0.58 & DES J010547.21--260321.9 ? & G & p & (1) & 23.52r & FR\,II relic \\ \\

J0106--2124  & 2.20 & 0.097 & 0.24 & 2MASX J01065914--2124484 & G & p & (1,8,11) & 16.19r & FR\,I/II, WAT relic \\
J0106--2247  & 1.27 & 0.42 & 0.42 & DES J010648.27--224712.2 & G & p & (1,8) & 18.40r & FR\,I, WAT? \\
J0106--2349  & 2.07 & 0.34 & 0.60 & DES J010606.86--234946.3 & G & p & (1,8) & 19.21r & FR\,II remnant \\
J0106--2539  & 3.03 & 0.90 & 1.42 & DES J010653.38--253904.0 & G & p & (1) & 21.42r & FR\,II relic \\
J0106--2559  & 4.79 & 0.07 & 0.39 & 2MASX J01061008--2559002 & G & p & (1) & 15.66r & FR\,II  \\
J0107--2245  & 2.05 & 1.04 & 0.99 & DESI J016.8794--22.7537 ? & ? & p & (1) & 25.34r & FR\,II bent, uncertain \\ 
J0107--2309  & 1.75 & 0.90 & 0.82 & DES J010714.99--230911.2 & QSO & s & (4) & 19.02r & FR\,II \\ 
{\bf J0107--2347}  & {\bf 13.8} & 0.312 & {\bf 3.79} & 2MASX J01072137--2347346 & G & p & (1) & 17.75r & {\bf FR\,II remnant, DDRG} \\ 
J0107--2510  & 1.43 & 1.0 & 0.69 & DES J010759.94--251019.3 & G & p & (1) & 22.85r & FR\,II \\
J0107--2541  & 0.85 & 0.58 & 0.34 & DES J010743.42--254108.8 & G & p & (1) & 20.24r & amorphous halo, end-on? \\ \\

J0107--2604  & 2.45 & 0.373 & 0.76 & DES J010703.32--260403.8 & G & p & (1) & 18.70r & FR\,II remnant \\
J0108--2442C & 3.15 & 0.85 & 1.45 & DES J010835.55--244227.3 & G & p & (1) & 22.39r & FR\,II remnant  \\
J0108--2454  & 1.26 & 0.23 & 0.28 & DES J010805.37--245447.4 & G & p & (1) & 18.67r & FR\,II remnant \\
J0108--2521  & 1.53 & 0.51 & 0.57 & DES J010837.82--252109.6 & G & p & (1) & 20.04r & FR\,II bent, WAT? \\
J0108--2600  & 2.4 & 0.54 & 0.91 & DES J010828.00--260047.9 & G & p & (1) & 19.75r & FR\,II relic \\
J0109--2426  & 1.45 & 0.56 & 0.56 & DES J010955.25--242605.2 & G & p & (1) & 20.54r & FR\,II remnant \\
J0110--2356  & 1.67 & 0.43 & 0.56 & DES J011029.06--235618.0 & G & p & (1) & 20.12r & FR\,I/II, WAT, precess. \\
J0110--2410  & 0.8 & 0.64 & 0.33 & DES J011022.64-241021.3 & G & p & (1) & 20.65r & FR\,II \\
J0110--2413  & $>$1.8 & 0.31 & $>$0.49 & DES J011016.69--241327.1 & G & p & (1,8) & 18.72r & FR\,II, complex \\ 
J0110--2430  & 0.53 & 0.40 & 0.17 & DES J011034.16--243030.5 & G & p & (1,8) & 18.82r & FR\,II \\ \\

J0111--2348  & 0.7 & 1.07 & 0.34 & DES J011100.79--234850.1 & G & p & (1) & 23.66r & FR\,II \\
J0111--2437  & 1.91 & 0.19 & 0.36 & 2MASX J01110877--2437384 & G & p & (1,8,11) & 16.68r & FR\,II \\ 
J0112--2358  & 2.93 & 1.14 & 1.45 & DES J011234.81--235833.1 & G & p & (1) & 23.62 & FR\,II \\
J0112--2453  & 1.07 & 1.03 & 0.52 & DES J011219.01--245321.0 & G & p & (1) & 22.43r & FR\,II relic \\
J0113--2543  & 1.15 & 0.2492 & 0.27 & DES J011306.70--254333.2 & G & s & (2) & 17.75r & FR\,II relic \\
\hline
\end{tabular}
\end{table*}

\end{document}